\documentclass[journal,article,submit,moreauthors,dvipdfm,12pt,a4paper]{mdpi} 

\usepackage{graphicx,bm}
\usepackage{amsmath}
\usepackage{amssymb}
\usepackage{amsfonts}

\newcommand{\be}{\begin{equation}}
\newcommand{\ee}{\end{equation}}
\newcommand{\bea}{\begin{eqnarray}}
\newcommand{\eea}{\end{eqnarray}}
\newcommand{\beaa}{\begin{eqnarray*}}
\newcommand{\eeaa}{\end{eqnarray*}}

\newcommand{\nn}{\nonumber \\}
\newcommand{\e}{\mathrm{e}}

\lastpage{x}
\doinum{10.3390/------}
\pubvolume{xx}
\pubyear{2012}
\history{Received: xx / Accepted: xx / Published: xx}

\Title{Stability of accelerating cosmology in two scalar-tensor theory: \\
Little Rip versus de Sitter}

\Author{Yusaku Ito$^1$, Shin'ichi Nojiri$^{1,2,\star}$,
and Sergei D. Odintsov$^{3,4}$
}

\address{%
$^1$ Department of Physics, Nagoya University, Nagoya
464-8602, Japan \\
$^2$ Kobayashi-Maskawa Institute for the Origin of Particles and
the Universe, Nagoya University, Nagoya 464-8602, Japan \\
$^3$Instituci\`{o} Catalana de Recerca i Estudis Avan\c{c}ats
(ICREA) and Institut de Ciencies de l'Espai (IEEC-CSIC), Campus
UAB, Facultat de Ciencies, Torre C5-Par-2a pl, E-08193 Bellaterra
(Barcelona), Spain \\
$^4$ Tomsk State Pedagogical
University, Tomsk, Russia}

\corres{nojiri@phys.nagoya-u.ac.jp. Tel. and FAX +81-45-789-4537}

\abstract{
We develop the general reconstruction scheme in two scalar model. The
quintom-like theory which may describe (different) non-singular
Little Rip or
de Sitter cosmology is reconstructed. The number of scalar phantom 
dark energy models (with Little Rip cosmology or 
asymptotically de Sitter evolution) is presented. 
Stability issue of such 
dark energy cosmologies as well as the flow to fixed points is studied.
The stability of
Little Rip universe which leads to dissolution of bound objects
sometime in future indicates that no classical transition to de Sitter
space occurs. The possibility of unification of inflation with Little Rip 
dark energy in two scalar theory is briefly mentioned.}



\begin{document}

\section{Introduction \label{I}}

The observation of the cosmic microwave background radiation (CMBR) shows that
the present universe is spatially flat 
\cite{de Bernardis:2000gy,Hanany:2000qf}.
Hence, the starting point  for current cosmology is
spatially flat Friedman-Lemaitre-Robertson-Walker (FLRW) universe,
whose metric is given by
\be
\label{JGRG14}
ds^2 = - dt^2 + a(t)^2 \sum_{i=1,2,3} \left(dx^i\right)^2\, .
\ee
Here $a$ is called  the scale factor. 
The FLRW equations in the Einstein gravity coupled with perfect 
fluid are well-known to be:
\be
\label{JGRG11}
\rho =\frac{3}{\kappa^2}H^2 \, ,\quad
p = - \frac{1}{\kappa^2}\left(3H^2 + 2\dot H\right)\, .
\ee
Here the Hubble rate $H$ is defined by $H\equiv \dot a/a$.
In (\ref{JGRG11}),  $\rho$ and $p$ are the energy density and the pressure of 
the perfect fluid.
The equation of state (EoS) parameter $w$ is defined by the ratio
of the pressure $p$ and the energy density $w\equiv p/\rho$.

On the other hand, the observation of the type Ia supernovae shows that the 
expansion of the present universe is
accelerating \cite{Perlmutter:1998np,Perlmutter:1997zf,Riess:1998cb}.
Before  the accelerating expansion of the present universe was established,
there were only two scenario about the future of the universe,
that is, eternal expansion and big crunch, where the universe  shrinks to
a point.
The discovery of the accelerating expansion indicates the possibility of 
another fate of the future universe, that is, the universe might evolve 
to finite-time future singularity.
What generates the accelerating expansion of the universe is called dark
energy (for recent review, see 
\cite{Silvestri:2009hh,Li:2011sd,Caldwell:2009ix,
Copeland:2006wr,Clifton:2011jh}, which is usually 
considered to be a perfect fluid with negative equation of state. 
The future of the universe is mainly governed by the equation of state (EoS) 
parameter $w_\mathrm{DE}$ of the dark energy 
$w_\mathrm{DE}\equiv p_\mathrm{DE} / \rho_\mathrm{DE}$,
where $p_\mathrm{DE}$ and $\rho_\mathrm{DE}$ are the pressure and the energy 
density of the dark energy.
The accelerating expansion can be generated if $w_\mathrm{DE} <-1/3$.
The observational data indicate that $w_\mathrm{DE}$ is close to $-1$.
In fact, we obtain $w=-1.068^{+0.080}_{-0.082}$ from the Supernova Legacy Survey 
\cite{Conley:2011ku,Sullivan:2011kv}, $w=-1.10\pm 0.14$ by combining 
the Cosmic Microwave Radiation (CMB) \cite{Komatsu:2010fb,Dunkley:2010ge} with 
the Sloan Digital Sky Survey \cite{Reid:2009xm}, and $w=-1.01^{+0.30}_{-0.40}$ 
by using the integrated Sachs-Wolfe effect from the CMB 
\cite{Giannantonio:2008zi,Ho:2008bz}. 
If the $w_\mathrm{DE}$ is exactly $-1$, the present universe is described by 
the $\Lambda$CDM model, where the cosmological term generates 
the accelerating expansion
and the universe  evolves to the asymptotic de Sitter space-time.
The dark energy with $-1<w_\mathrm{DE}<-1/3$ is called as quintessence and that
with $w_\mathrm{DE}<-1$ as phantom. If the dark energy is the phantom, 
the future universe usually evolves to a
finite-time future singularity called  Big Rip \cite{astro-ph/0302506}, where
the scale factor $a$ of the universe will diverge in the finite future.
Even if the dark energy is quintessence, soft singularity might be generated in 
the finite-time future.
The classification of the finite-time future singularities is given in
\cite{hep-th/0501025}, as follows:
\begin{itemize}
\item Type I (``Big Rip'') : For $t \to t_s$, $a \to \infty$,
$\rho_\mathrm{DE} \to \infty$ and $\left|p_\mathrm{DE}\right| \to \infty$.
This also includes the case of $\rho_\mathrm{DE}$, $p_\mathrm{DE}$ being
finite at $t_s$.
\item Type II (``sudden'') \cite{Barrow:2004xh,Barrow:2004hk}:
For $t \to t_s$, $a \to a_s$, $\rho_\mathrm{DE} \to \rho_s$ and
$\left|p_\mathrm{DE}\right| \to \infty$.
\item Type III : For $t \to t_s$, $a \to a_s$,
$\rho_\mathrm{DE} \to \infty$ and
$\left|p_\mathrm{DE}\right| \to \infty$.
\item Type IV : For $t \to t_s$, $a \to a_s$,
$\rho_\mathrm{DE} \to 0$, $\left|p_\mathrm{DE}\right| \to 0$ and higher
derivatives of $H$ diverge.
This also includes the case in which $p_\mathrm{DE}$
($\rho_\mathrm{DE}$) or both of $p_\mathrm{DE}$ and $\rho_\mathrm{DE}$
tend to some finite values, whereas higher derivatives of $H$ diverge.
\end{itemize}
The natural prescription to cure finite-time future singularity may be found
in frames of number of viable models of modified gravity \cite{arXiv:1011.0544}.
However, it is still very interesting to understand if the solution of
singularity problem may be found within fluid dark energy.

Let us concentrate now on phantom era which normally lead to Big Rip 
singularity in finite future. Of course, phantom era may be transient and due to 
coupling of dark energy with dark matter, or due to complicated dynamics of 
dark energy itself, etc it may end up in asymptotically de Sitter space or even 
in deceleration regime.
The Big Rip singularity is usually generated if $w<-1$. Let us assume that EoS 
parameter is always less than $-1$. Even if $w<-1$, if $w$ approaches
to $-1$ sufficiently rapidly, the singularity is not always generated.
First of all, transient phantom era is possible.
Moreover, one construct the phantom models where $w$ asymptotically tends to 
$-1$ so that the universe ends up in asymptotically de Sitter space.
Evidently, there is no future singularity in such a case.
Recently, new scenario to avoid future singularity has been proposed in
\cite{Frampton:2011sp} (for further development see
Refs.~\cite{Frampton:2011rh,Nojiri:2011kd,arXiv:1107.4642,
arXiv:1103.2480,arXiv:1111.2454} and a similar scenario to 
avoid the singularity was also considered in  \cite{Capozziello:2009hc}).
In this scenario, since $w\to -1$ asymptotically ($w$ always remains to be less 
than $-1$), the finite-time singularity is avoided.
In other words, the singularity is removed to infinite future.
Even in such non-singular cosmology, if $H$ goes to infinity when $t\to \infty$,
it was found that there might occur the dissolution
of bound objects sometime in future, similarly to Big Rip singularity.
That is why the scenario was called Little Rip cosmology. 
The phantom scalar models 
to describe Little Rip were introduced in Ref.~\cite{Frampton:2011rh}.
Finally, there is possibility of pseudo-rip scenario \cite{Frampton:2011aa} 
which lies between Little Rip and cosmological constant cosmology with Hubble 
rate tending to constant at infinite future.
It is known that in one scalar models, the large instability occurs when
crossing the phantom-divide (or cosmological constant border). In order to
understand Little Rip cosmology better, as well as its relation with
asymptotically de Sitter
universe and possible transitions between these two spaces, the more
realistic description of Little Rip may be necessary.
In this paper, we will also show the possibility of ``Big Crush'' in
(\ref{lc1}), where $H$ and $a$ are finite but all structures are crushed
by the infinitesimally large inertial force. The Big Crush can be generated
by the Type II singularity.

In the present paper we develop phantom cosmology description in terms of 
two-scalar tensor theory which represents kind of quintom model
\cite{arXiv:astro-ph/0404224,hep-th/0405034}
(for review, see \cite{arXiv:0909.2776} and for generalizations,
see \cite{Nojiri:2005pu,arXiv:0811.3643}).
Non-singular phantom cosmology in frames of such theory is described and 
stability of different cosmological scenarios is investigated.

In Section \ref{II}, we review the one scalar model in order to clarify the 
problem of the infinite instability. After that, we consider a general 
formulation of reconstruction in two scalar model and investigate the 
stability of the solution.
In this formulation, we construct a model which has a stable cosmological
solution describing the phantom-divide crossing.
In Section \ref{III}, we reconstruct a model which
describes the cosmological solutions with and without Little Rip and
investigate the (in)stability of the solutions. The existence of the
solution describing de Sitter space-time and the stability of
the de Sitter solution when it exists as well as possible transition of
Little Rip cosmology to de Sitter one are investigated.
In Section \ref{IV}, we also consider the reconstruction of the two scalar 
model in terms of the e-foldings $N$ and investigate the flow of the solution
in terms of dimensionless variables, which give the fixed points for some 
solutions.
Some summary and outlook are given in Discussion section.

\section{Reconstruction of scalar model and (in)stability \label{II}}

Recent cosmological data seems to indicate that there occurred the crossing of
the phantom divide line in the near past (see, for example, 
\cite{Nesseris:2006er}),
that is, the equation of state (EoS) parameter $w$ crossed the line $w=-1$.
Then it is interesting to construct a model generating the phantom crossing.
Here we consider such a model using the scalar field.
For the model with one scalar, however, it is known that
the instability becomes very large
when crossing the cosmological constant line $w=-1$, corresponding to
the transition from the quintessence phase to the phantom phase.
In order to avoid this problem, one may consider 
two scalar model \cite{Nojiri:2005pu,Capozziello:2005tf}.

\subsection{One scalar model}

Before going to the two scalar model, we review on the reconstruction of the 
one scalar model and clarify the problem of the instability based on
\cite{Nojiri:2005pu,Capozziello:2005tf}.

We start with the following action:
\be
\label{ma7}
S=\int d^4 x \sqrt{-g}\left\{
\frac{1}{2\kappa^2}R - \frac{1}{2}\omega(\phi)\partial_\mu \phi
\partial^\mu\phi - V(\phi) + L_\mathrm{matter} \right\}\, .
\ee
Here, $\omega(\phi)$ and $V(\phi)$ are functions of the scalar field $\phi$.
The function $\omega(\phi)$ is not relevant and can be absorbed into the
redefinition of the scalar field $\phi$. In fact, if one redefines
the scalar field $\phi$ by
\be
\label{ma13}
\varphi \equiv \int^\phi d\phi \sqrt{\left|\omega(\phi)\right|} \, ,
\ee
the kinetic term of the scalar field in the action (\ref{ma7}) has the
following form:
\be
\label{ma13b}
 - \omega(\phi) \partial_\mu \phi \partial^\mu\phi
= \left\{ \begin{array}{ll}
 - \partial_\mu \varphi \partial^\mu\varphi &
\mbox{when $\omega(\phi) > 0$} \\
\partial_\mu \varphi \partial^\mu\varphi & \mbox{when $\omega(\phi) < 0$}
\end{array} \right. \, .
\ee
The case of $\omega(\phi) > 0$ corresponds to the quintessence or
non-phantom scalar field, but the case of $\omega(\phi) < 0$ corresponds
to the phantom scalar.
Although $\omega(\phi)$ can be absorbed into the redefinition of the
scalar field, we keep $\omega(\phi)$ since the transition between the
quintessence and the phantom can be described by the change of the sign
of $\omega(\phi)$.

For the action (\ref{ma7}), the energy density and the pressure of the scalar 
field are given as follows:
\be
\label{ma8}
\rho = \frac{1}{2}\omega(\phi){\dot \phi}^2 + V(\phi)\, ,\quad
p = \frac{1}{2}\omega(\phi){\dot \phi}^2 - V(\phi)\, .
\ee
If the potential $V$ is positive, the EoS parameter $w$ is greater than $-1$
if $\omega(\phi)$ is positive but $w$ is less than $-1$
if $\omega(\phi)$ is negative.
Then the transition between the quintessence and the phantom can be 
described by the change of the sign of $\omega(\phi)$.
By using the FLRW equation (\ref{JGRG11}), the EoS parameter $w$ can be
expressed in the following form
\be
\label{i2}
w = - 1 - \frac{2\dot H}{3H^2}\, .
\ee
Then in the quintessence phase, where $1>1/3 w >-1$, we find $\dot H<0$ and
in the phantom phase, where $w<-1$, $\dot H>0$. Then on the point of the
transition between the quintessence and the phantom, $\dot H$ vanishes.

In order to consider and explain the cosmological reconstruction in terms of 
one scalar model, we rewrite the FLRW equation (\ref{JGRG11}) with the
expressions (\ref{ma8}) as follows:
\be
\label{ma9}
\omega(\phi) {\dot \phi}^2 = - \frac{2}{\kappa^2}\dot H\, ,\quad
V(\phi)=\frac{1}{\kappa^2}\left(3H^2 + \dot H\right)\, .
\ee
Assuming $\omega(\phi)$ and $V(\phi)$ are given by a single function
$f(\phi)$, as follows,
\be
\label{ma10}
\omega(\phi)=- \frac{2}{\kappa^2}f'(\phi)\, ,\quad
V(\phi)=\frac{1}{\kappa^2}\left(3f(\phi)^2 + f'(\phi)\right)\, ,
\ee
the exact solution of the FLRW equations
(when we neglect the contribution from the matter)
has the following form:
\be
\label{ma11}
\phi=t\, ,\quad H=f(t)\, .
\ee
It can be confirmed that the equation given by the variation over $\phi$
\be
\label{ma12}
0=\omega(\phi)\ddot \phi + \frac{1}{2}\omega'(\phi){\dot\phi}^2
+ 3H\omega(\phi)\dot\phi + V'(\phi)\, ,
\ee
is also satisfied by the solution (\ref{ma11}).
Then, the arbitrary universe evolution expressed by $H=f(t)$ can be
realized by an appropriate choice of $\omega(\phi)$ and $V(\phi)$.
In other words, by defining the particular type of universe evolution,
the corresponding scalar-Einstein gravity may be found.

As mentioned in (\ref{ma13}) and (\ref{ma13b}), $\omega(\phi)$ can be
absorbed into the redefinition of the scalar field $\phi$.
By keeping $\omega(\phi)$, however, we can construct a model that exhibits
a smooth transition between the non-phantom phase and the phantom phase.

We now show that there occurs the very large instability when
crossing the cosmological constant line $w=-1$.
For this purpose, by introducing the new variables $X_\phi$ and $Y$
as follows
\be
\label{R1}
X \equiv \dot \phi\, ,\quad Y \equiv \frac{f(\phi)}{H}\, ,
\ee
we rewrite the FLRW equation (\ref{JGRG11}) with (\ref{ma8})
and the field equation (\ref{ma12}) as \be
\label{R2}
\frac{d X_\phi}{d_N} = - \frac{f''(\phi) \left( X_\phi^2 - 1 \right)}{2 f' 
(\phi) H}
 - 3 \left( X_\phi - Y \right) \, ,\quad
\frac{d Y}{dN} = \frac{f'(\phi) \left( 1 - X_\phi Y \right) X_\phi}{H^2}\, .
\ee
Here $N$ is called as e-foldings and $d/dN\equiv H^{-1}d/dt$.
Since we have $X_\phi=Y=1$ for the solution (\ref{ma11}), we now consider the 
following perturbation:
\be
\label{R3}
X_\phi = 1 + \delta X_\phi\, , \quad Y = 1 + \delta Y\,
\ee
Then
\be
\label{R4}
\frac{d}{dN} \left( \begin{array}{c}
\delta X_\phi \\
\delta Y
\end{array} \right)
= \left( \begin{array}{cc}
 - \frac{\ddot H}{\dot H H} - 3 & 3 \\
 - \frac{\dot H}{H^2} & - \frac{\dot H}{H^2}
\end{array} \right) \left( \begin{array}{c}
\delta X_\phi \\
\delta Y
\end{array} \right)\, .
\ee
Here, the solution (\ref{ma11}) is used.
The eigenvalues $M_\pm$ of the matrix are given by
\be
\label{R5}
M_\pm = \frac{1}{2} \left\{ - \left( \frac{\ddot H}{\dot H H}
+ \frac{\dot H}{H^2} + 3 \right) \pm \sqrt{ \left( \frac{\ddot H}{\dot H H}
+ \frac{\dot H}{H^2} + 3 \right)^2 - \frac{4\ddot H}{H^3} 
 - \frac{12 \dot H}{H^2}} \right\}\, .
\ee
In order that the solution (\ref{ma11}) could be stable, all the eigenvalues
$M_\pm$ must be negative. One now considers the region near the transition 
between the quintessence phase and the phantom phase, where $\dot H\sim 0$.
When we consider the transition from the quintessence phase, where $\dot H<0$,
to the phantom phase, we find $\ddot H > 0$.
On the other hand, when we consider the transition from the phantom phase,
where $\dot H>0$, to the quintessence phase, we find $\ddot H < 0$.
Then for both transitions, one finds $\ddot H / \dot H H$ is large
and negative. The eigenvalues $M_\pm$ are given by
\be
\label{R6}
M_+ \sim - \frac{\ddot H}{\dot H H}\, ,\quad M_- \sim 0\, .
\ee
Then $M_+$ is positive and diverges on the point of the transition $\dot H=0$.
Hence, the solution describing the transition is always unstable and the 
instability diverges at the transition point and therefore  the transition 
is prohibited in the one scalar model.
This instability was first observed in \cite{Vikman:2004dc} 
and in \cite{Xia:2007km}, it was shown that a generic $k$-essence-like scalar 
field model minimally coupled to the Einstein gravity is not allowed to give the 
transition since its perturbation modes are divergent at that moment. 

\subsection{Two scalar model}

As we have shown, in the case of the one scalar model, the instability becomes 
infinite at the crossing $w=-1$ point. Let us show that such
a divergence of the instability does not occur for model with two scalars.

We now consider the following two scalar model
\be
\label{A1}
S=\int d^4 x \sqrt{-g}\left\{\frac{1}{2\kappa^2}R
 - \frac{1}{2}\omega(\phi)\partial_\mu \phi
\partial^\mu \phi - \frac{1}{2}\eta(\chi)\partial_\mu \chi
\partial^\mu \chi - V(\phi,\chi)
\right\}\ .
\ee
Here $V(\phi,\chi)$ is the potential and $\omega(\phi)$ and $\eta(\chi)$ are 
functions of the scalar field $\phi$ and $\chi$, respectively.
We should note that the functions $\omega(\phi)$ and $\eta(\chi)$ can be 
absorbed into the redefinition of the scalar fields as in case of one 
scalar model (\ref{ma13}) but for the later convenience, we keep these functions. 
By keeping these functions, one can perform
the reconstruction of the model  explicitly as we will see below.

For the action (\ref{A1}), the FLRW equations give
\be
\label{A2}
\omega(\phi) {\dot \phi}^2 + \eta(\chi) {\dot \chi}^2 
= - \frac{2}{\kappa^2}\dot H\ ,\quad
V(\phi,\chi)=\frac{1}{\kappa^2}\left(3H^2 + \dot H\right)\ .
\ee
Then if
\be
\label{A3}
\omega(t) + \eta(t)=- \frac{2}{\kappa^2}f'(t)\ ,\quad
V(t,t)=\frac{1}{\kappa^2}\left(3f(t)^2 + f'(t)\right)\ ,
\ee
the explicit solution follows
\be
\label{A4}
\phi=\chi=t\ ,\quad H=f(t)\ .
\ee
One may choose that $\omega$ should be always positive and $\eta$ be
always negative, for example
\begin{align}
\label{A5}
\omega(\phi)&=-\frac{2}{\kappa^2}\left\{f'(\phi) - \sqrt{\alpha(\phi)^2 +
f'(\phi)^2}\right\}>0\ ,\nn
\eta(\chi)&=-\frac{2}{\kappa^2}\sqrt{\alpha(\chi)^2 + f'(\chi)^2}<0\ .
\end{align}
Here $\alpha$ is an arbitrary real function.
We now define a new function $\tilde f(\phi,\chi)$ by
\be
\label{A6}
\tilde f(\phi,\chi)\equiv - \frac{\kappa^2}{2}\left(\int d\phi \omega(\phi)
+ \int d\chi \eta(\chi)\right)\ ,
\ee
which gives $\tilde f(t,t)=f(t)$.
If $V(\phi,\chi)$ is given by using $\tilde f(\phi,\chi)$ as
\be
\label{A8}
V(\phi,\chi)=\frac{1}{\kappa^2}\left(3{\tilde f(\phi,\chi)}^2
+ \frac{\partial \tilde f(\phi,\chi)}{\partial \phi}
+ \frac{\partial \tilde f(\phi,\chi)}{\partial \chi} \right)\ ,
\ee
the FLRW and the following scalar field equations are also satisfied:
\begin{align}
\label{A9}
0&=\omega(\phi)\ddot\phi + \frac{1}{2}\omega'(\phi) {\dot \phi}^2
+ 3H\omega(\phi)\dot\phi + \frac{\partial
\tilde V(\phi,\chi)}{\partial \phi}\ ,\nn
0&=\eta(\chi)\ddot\chi + \frac{1}{2}\eta'(\chi) {\dot \chi}^2
+ 3H\eta(\chi)\dot\chi + \frac{\partial \tilde V(\phi,\chi)}{\partial \chi}\, .
\end{align}
Choosing $\omega(\phi)$, $\eta(\chi)$, and $V(\phi,\chi)$ in the action
(\ref{A1}) as in (\ref{A5}) and (\ref{A8}), we obtain a model which has a 
solution (\ref{A4}) as an exact solution.

In case of the one scalar model, the instability becomes infinite
at the crossing $w=-1$ point. One may expect that such
a divergence of the instability does not occur for model with two scalars.

By introducing the new quantities, $X_\phi$, $X_\chi$, and ${\tilde Y}$ as
\be
\label{A10}
X_\phi \equiv \dot \phi\ ,\quad X_\chi \equiv \dot \chi\ ,\quad
{\tilde Y}\equiv \frac{\tilde f(\phi,\chi)}{H} \ ,
\ee
the FLRW equations and the scalar field equations (\ref{A9}) are rewritten as
\begin{align}
\label{A11}
\frac{dX_\phi}{dN}&= - \frac{\omega'(\phi)}{2H \omega(\phi)}
\left(X_\phi^2 - 1\right) - 3(X_\phi-{\tilde Y})\ ,\nn
\frac{dX_\chi}{dN}&= - \frac{\eta'(\chi)}{2H \eta(\chi)}
\left(X_\chi^2 - 1\right) - 3(X_\chi-{\tilde Y})\ ,\nn
\frac{d{\tilde Y}}{dN}&
= \frac{3 X_\phi X_\chi \left( 1 - {\tilde Y}^2 \right)}{X_\phi + X_\chi}
+ \frac{\dot H}{H^2} \frac{X_\phi X_\chi + 1 - {\tilde Y}
\left( X_\phi + X_\chi\right)}{X_\phi + X_\chi} \, .
\end{align}
In the solution (\ref{A4}), we find $X_\phi=X_\chi={\tilde Y}=1$.
The following perturbation may be considered:
$X_\phi=1+\delta X_\phi$, $X_\chi=1 + \delta X_\chi$, and
${\tilde Y}=1 + \delta {\tilde Y}$.
Hence
\be
\label{A13}
\frac{d}{dN}\left(\begin{array}{c}
\delta X_\phi \\
\delta X_\chi \\
\delta {\tilde Y}
\end{array}\right)
= M \left(\begin{array}{c}
\delta X_\phi \\
\delta X_\chi \\
\delta {\tilde Y}
\end{array}\right)
\ ,\quad
M\equiv \left(\begin{array}{ccc}
 - \frac{\omega'(\phi)}{H\omega(\phi)} - 3 & 0 & 3 \\
0 & - \frac{\eta'(\chi)}{H\eta(\chi)} - 3 & 3 \\
0 & 0 & - 3 - \frac{\dot H}{H^2}
\end{array}\right)\, .
\ee
The eigenvalues of the matrix $M$ are given by
\be
\label{AAA1}
M_\phi \equiv - \frac{\omega'(\phi)}{H\omega(\phi)} - 3 \, ,\quad
M_\chi \equiv - \frac{\eta'(\chi)}{H\eta(\chi)} - 3 \, ,\quad
M_{\tilde Y} \equiv - 3 - \frac{\dot H}{H^2} \, .
\ee
The eigenvalues (\ref{AAA1}) for the two scalar model are clearly finite.
Hence, the instability, if any, could be finite and by choosing $\alpha$
in (\ref{A5}) properly, the instability can be removed, in general.
In fact, right on the transition point where $\dot H=f'(t)=0$
and therefore $f'(\phi)=f'(\chi)=0$, for the choice in (\ref{A5})
with constant $\alpha$, $\alpha(\phi)=\alpha(\chi)=\alpha>0$, we find
$\omega(\phi)=-\eta(\chi) = 2\alpha/\kappa^2$,
$\omega'(\phi)=-2\ddot H/\kappa^2$, and $\eta'(\chi)=0$.
Then the eigenvalues (\ref{AAA1}) reduce to
\be
\label{AA2}
M_\phi = \frac{\ddot H}{\alpha H} - 3\, ,\quad
M_\chi = M_{\tilde Y} = - 3 \, .
\ee
Then as long as $\frac{\ddot H}{\alpha H} < 3$, all the eigenvalues are
negative and therefore the solution (\ref{A4}) is stable.

We should note that the solution (\ref{A4}) is merely one of the solution 
of the FLRW equations (\ref{A2}) although the solution (\ref{A4}) is exact 
and the field equations (\ref{A9}) and there are other 
kind of solutions. If the solution (\ref{A4}) is stable, however, the solution becomes 
an attractor solution and there should be a class of solutions which approaches to the 
solution (\ref{A4}) asymptotically. 

Hence, we gave general formulation of reconstruction in two scalar model and
investigated the stability of the solution. Using this formulation, one can
construct a model which has
a stable cosmological solution corresponding to the phantom-divide crossing.

\section{Reconstruction of Little Rip cosmology \label{III}}

As the universe expands, the relative acceleration between two points separated 
by a distance $l$ is given by $l \ddot a/a$, where $a$ is the scale factor. 
If there is a particle with mass $m$ at each of the points, an observer at one 
of the masses will measure an inertial force on the other mass of
\be
\label{i1}
F_\mathrm{iner}=m l \ddot a/a = m l \left( \dot H + H^2 \right)\, .
\ee
Let us assume the the two particles are bound by a constant force $F_0$. If 
$F_\mathrm{iner}$ is positive and greater
than $F_0$, the two particles become unbound. This is the ``rip'' produced by 
the accelerating expansion.
Note that equation (\ref{i1}) shows that a rip always occurs when either $H$ 
diverges or $\dot H$
diverges (assuming $\dot H > 0$). The first case corresponds to
a ``Big Rip'' \cite{Caldwell:1999ew}, while
if $H$ is finite, but $\dot H$ diverges with $\dot H > 0$,
we have a Type II or ``sudden future'' singularity
\cite{Barrow:2004xh,Barrow:2004hk}, which also leads to a rip.
Even if $H$ or $\dot H$ goes to infinity at the infinite future, the inertial 
force becomes larger and larger,
and any bound object is ripped, which is called ``Little Rip''.

The acceleration by the gravitational force between the sun and the earth is
given by $a_\mathrm{g} = l \omega_A^2$.
Here $l$ is the distance between the sun and the earth and $\omega_A$ is the
angular speed
$\omega_A =2\pi/\left(1\, \mathrm{year}\right) = 1.99\times 10^{-6}\, 
\mathrm{s}^{-1}$.
If the acceleration $a_\mathrm{e}$ of the inertial force by the expansion
(\ref{i1}) exceeds $a_\mathrm{g}$,
there occurs the rip between the earth and the sun, that is,
\be
\label{Acc}
a_\mathrm{e} = l \left( \dot H + H^2 \right) \sim l H^2 > a_\mathrm{g}\, .
\ee

An interesting case occurs when $H$ is finite and $\dot H$ diverges but is 
negative. In this case,
even though the universe is expanding, all structures are crushed rather than 
ripped. An example is given by
\be
\label{lc1}
H = H_0^{(0)} + H_1^{(0)} \left(t_c - t\right)^\alpha\, .
\ee
Here $H_0^{(0)}$ and $H_1^{(0)}$ are positive constants and $\alpha$ is a 
constant with $0<\alpha<1$.

In this section, by using the formulation of the previous section,
we construct a model which may generate a Little Rip cosmology
\cite{Frampton:2011sp}.

\subsection{A model of Little Rip cosmology}

As an example of the model generating the Little Rip, we consider the following 
Hubble rate
\be
\label{lr1}
H = H_0^{(\mathrm{I})} \e^{\lambda t}\, ,
\ee
Here $H_0^{(\mathrm{I})}$ and $\lambda$ are positive constants.
Eq.~(\ref{lr1}) shows that there is no curvature singularity for finite $t$.

When we ignore the contribution from matter, the equation of state (EoS)
parameter $w$ of the dark energy can be expressed in terms of the Hubble rate 
$H$
as in (\ref{i2}). Then if $\dot H>0$,  $w<-1$.
By using Eq.~(\ref{i2}), one finds
\be
\label{i4}
w = -1 - \frac{2\lambda}{3H_0^{(\mathrm{I})}} \e^{-\lambda t}\, ,
\ee
and therefore $w<-1$ and $w\to -1$ when $t\to +\infty$, and
$w$ is always less than $-1$ when $\dot H$ is positive.
The parameter $A$ in \cite{Frampton:2011sp} corresponds to $2\lambda/\sqrt{3}$
in (\ref{lr1}) and is bounded as
\be
\label{RRR1}
2.37\times 10^{-3}\, \mathrm{Gyr}^{-1} \leq \lambda \leq
8.37\times 10^{-3}\, \mathrm{Gyr}^{-1}\, ,
\ee by the results of the Supernova Cosmology Project
\cite{Amanullah:2010vv}.

In the model (\ref{lr1}), $H$ is always finite but increases exponentially,
which generates the strong inertial force. 
The inertial force becomes larger and larger and any bound object is ripped. 
This phenomenon is a ``Little Rip'' \cite{Frampton:2011sp}.

Now Little Rip cosmology may be realized in two-scalar model.
By choosing $\alpha$ in (\ref{A5}) as
\be
\label{lr2}
\alpha (t) = \alpha_0^{(\mathrm{I})} \e^{\lambda t}\, ,
\ee
with a constant $\alpha_0^{(\mathrm{I})}$,
we find $\omega(\phi)$ and $\eta(\chi)$ in (\ref{A5}) as follows
\be
\label{lr3}
\omega (\phi) = \frac{2}{\kappa^2} \left( 
\sqrt{\left(\alpha_0^{(\mathrm{I})}\right)^2
+ \lambda^2 \left(H_0^{(\mathrm{I})}\right)^2} - \lambda H_0^{(\mathrm{I})} 
\right) \e^{\lambda \phi}\, ,
\quad \eta (\chi) = - \frac{2}{\kappa^2} 
\sqrt{\left(\alpha_0^{(\mathrm{I})}\right)^2
+ \lambda^2 \left(H_0^{(\mathrm{I})}\right)^2}\,
\e^{\lambda \chi} \, .
\ee
Using (\ref{A6}), we obtain
\be
\label{lr4}
\tilde f(\phi,\chi) = \frac{1}{\lambda} \left\{
 - \left( \sqrt{\left(\alpha_0^{(\mathrm{I})}\right)^2
+ \lambda^2 \left(H_0^{(\mathrm{I})}\right)^2}
 - \lambda H_0^{(\mathrm{I})} \right)\e^{\lambda \phi}
+ \sqrt{\left(\alpha_0^{(\mathrm{I})}\right)^2 + \lambda^2
\left(H_0^{(\mathrm{I})}\right)^2}\,\e^{\lambda \chi}\right\}\, ,
\ee
and the potential in (\ref{A8}) is given by
\begin{align}
\label{lr5}
V(\phi,\chi) =& \frac{1}{\kappa^2} \left[ \frac{3}{\lambda^2}
\left\{ - \left( \sqrt{\left(\alpha_0^{(\mathrm{I})}\right)^2
+ \lambda^2 \left(H_0^{(\mathrm{I})}\right)^2} - \lambda H_0^{(\mathrm{I})} 
\right)
\e^{\lambda \phi} + \sqrt{\left(\alpha_0^{(\mathrm{I})}\right)^2
+ \lambda^2 \left(H_0^{(\mathrm{I})}\right)^2}\,\e^{\lambda \chi}\right\}^2 
\right. \nn
& \left. - \left( \sqrt{\left(\alpha_0^{(\mathrm{I})}\right)^2
+ \lambda^2 \left(H_0^{(\mathrm{I})}\right)^2} - \lambda H_0^{(\mathrm{I})} 
\right) \e^{\lambda \phi}
+ \sqrt{\left(\alpha_0^{(\mathrm{I})}\right)^2
+ \lambda^2 \left(H_0^{(\mathrm{I})}\right)^2}\,\e^{\lambda \chi}\right]\, ,
\end{align}
The parameter $\lambda$ is bounded as in (\ref{RRR1}).
If we choose time so that the present universe corresponds to $t=0$, we have
$H_0^{(\mathrm{I})} \sim 70 \mathrm{km}/\mathrm{s}\,\mathrm{Mpc}$.

If we define new fields $\theta$ and $\rho$ by
\begin{align}
\label{tr1}
\theta \equiv& \frac{2\sqrt{2}}{\kappa\lambda} \left( 
\sqrt{\left(\alpha_0^{(\mathrm{I})}\right)^2
+ \lambda^2 \left(H_0^{(\mathrm{I})}\right)^2} - \lambda H_0^{(\mathrm{I})} 
\right)^{\frac{1}{2}} \e^{\frac{\lambda \phi}{2}}\, , \nn
\rho \equiv& \frac{2\sqrt{2}}{\kappa \lambda} \left(
\sqrt{\left(\alpha_0^{(\mathrm{I})}\right)^2
+ \lambda^2 \left(H_0^{(\mathrm{I})}\right)^2} \right)^{|frac{1}{2}}
\e^{\frac{\lambda \chi}{2}} \, ,
\end{align}
we find
\begin{align}
\label{tr2}
& - \frac{1}{2}\omega(\phi)\partial_\mu \phi \partial^\mu\phi 
 - \frac{1}{2}\eta(\chi)\partial_\mu \chi\partial^\mu\chi - V(\phi,\chi) \nn
&= - \frac{1}{2}\partial_\mu \theta \partial^\mu\theta 
+ \frac{1}{2}\partial_\mu \rho \partial^\mu\rho
+ \frac{\lambda^2}{8}\left(\theta^2 - \rho^2\right)
- \frac{3}{64}\kappa^2 \lambda^2 \left(\theta^2 - \rho^2\right)^2\, . 
\end{align}
This Lagrangian density is nothing but the one of usual renormalizable scalar fields 
with mass term and the quartic potential as in the Higgs fields.  

For the model (\ref{lr1}) with (\ref{lr2}), the eigenvalues (\ref{AAA1}) are
given by corresponding expressions when $\lambda t \gg 1$,
\be
\label{lr5b}
M_\phi = M_\chi = M_{\tilde Y} = - 3 - 
\frac{\lambda}{H_0^{(\mathrm{I})}}\e^{-\lambda t}\, ,
\ee
which are negative. Therefore the solution is stable.

Let us consider the possibility that the universe could evolve to the 
de Sitter space-time.
In de Sitter space-time, since $H$ and $\dot H$ are finite and 
therefore the inertial force (\ref{i1}) is finite, there does not 
occur any rip.
Since the field $\phi$ is a usual canonical field, the minimum of the potential is stable. 
On the other hand, the field $\chi$ is non-canonical phantom field, the maximum of the potential 
is stable, which could be found from the field equation or the analogy of the classical equation 
of motion of the particle with negative mass. Therefore
in order for the solution corresponding to the de Sitter
space-time to exist, there should be an extremum in the potential and the
potential should be positive there.
If the extremum is local minimum with respect to $\phi$, which is the canonical
scalar and local maximum with
respect to $\chi$, which is non-canonical or phantom scalar, the solution is
stable.
For the potential (\ref{lr5}), there is an extremum when
\be
\label{stbl1}
 - \left( \sqrt{\left(\alpha_0^{(\mathrm{I})}\right)^2
+ \lambda^2 \left(H_0^{(\mathrm{I})}\right)^2}
 - \lambda H_0^{(\mathrm{I})} \right) \e^{\lambda \phi}
+ \sqrt{\left(\alpha_0^{(\mathrm{I})}\right)^2
+ \lambda^2 \left(H_0^{(\mathrm{I})}\right)^2}\,\e^{\lambda \chi}
= - \frac{\lambda^2}{6}\, ,
\ee
where the value of the potential $V(\phi,\chi)$ is given by
$V(\phi,\chi) = - \frac{\lambda^2}{12\kappa^2}$,
which is negative and therefore there does not exist the solution corresponding
to the de Sitter space-time.
Hence, the universe does not evolve into de Sitter space-time.


We now show that the (asymptotically) Little Rip solution (\ref{lr1}) is always
asymptotically stable.
For large $t$, one assumes the solution behaves as (\ref{lr1}).
Among three eigenvalues  (\ref{AAA1}), for the asymptotically Little Rip 
solution which is in the
phantom phase $\dot H>0$, the eigenvalue $M_{\tilde Y}$ is negative.
The eigenvalues and therefore the stability of the de Sitter solution 
depend on the choice of $\alpha(\chi)$ in (\ref{A5}). 
Then as an example, if we write $\alpha(\chi)$ as
$\alpha(\chi) = \lambda H_0^{(\mathrm{I})} q(\chi) \e^{\lambda \chi}$,
the eigenvalue $M_\chi$  (\ref{AAA1}) can be expressed as a function of
$t=\chi$ as follows,
\be
\label{ss2}
M_\chi = - \frac{q(t) \left( q'(t) + \lambda q(t) \right)
+ \lambda}{H_0^{(\mathrm{I})} \e^{\lambda t} \left( q(t)^2 + 1 \right)} - 3\, .
\ee
If $M_\chi$ could be positive, $q(t) q'(t)$ must be negative. Since $q(t)^2$ is
positive, $q(t)^2$ goes to a constant $q(t)^2 \to Q_0 \geq 0$. 
Then due to the factor $\e^{\lambda t}$ in the denominator of
the first term in (\ref{ss2}), the first term goes to small value for large $t$
and we find $M_\chi\to -3<0$. 
Therefore $M_\chi$ is asymptotically negative.
For the eigenvalue $M_\phi$, one gets
\be
\label{ss3}
M_\phi = - \frac{\lambda + \frac{q(t)q'(t)}{\sqrt{q(t)^2 + 1}
\left( \sqrt{q(t)^2 + 1} - 1 \right)}}{H_0^{(\mathrm{I})} 
\e^{\lambda t}} -3\, .
\ee
Then again in order that $M_\phi$ could be positive, $q(t) q'(t)$ must be
negative and therefore $q(t)^2$ goes to a constant $q(t)^2 \to Q_0 \geq 0$. 
Due to the factor $\e^{\lambda t}$ in the denominator of the first term 
in (\ref{ss3}), the first term goes to small value for large $t$
and $M_\phi\to -3<0$.
Therefore, all the eigenvalues are negative and the Little Rip solution is
asymptotically stable.


\subsection{Asymptotically de Sitter phantom model}

As another example, we consider the model where the EoS parameter $w$ is 
always less than $-1$ but there is no Big Rip nor Little Rip.
We consider the following the model:
\be
\label{i6}
H = H_0^{(\mathrm{II})} - H_1^{(\mathrm{II})} \e^{-\lambda t}\, .
\ee
Here $H_0^{(\mathrm{II})}$, $H_1^{(\mathrm{II})}$, and $\lambda$ are positive 
constants and we assume $H_0^{(\mathrm{II})}>H_1^{(\mathrm{II})}$ and $t>0$.
We use the same parameter $\lambda$ as in (\ref{lr1}) since it gives the same 
asymptotic behavior of the EoS parameter $w$ ($w+1 \sim \e^{-\lambda t}$) as 
we will see in (\ref{wn1}) (compare with (\ref{i4})).

Since the second term decreases when $t$ increases, the universe goes to
asymptotically de Sitter space-time.
Then from Eq.~(\ref{i2}), we find
\be
\label{i7}
w = -1 - \frac{2\lambda H_1^{(\mathrm{II})}
\e^{- \lambda t}}{3\left( H_0^{(\mathrm{II})} - H_1^{(\mathrm{II})} 
\e^{-\lambda t}
\right)^2} \, .
\ee
As in the previous example (\ref{lr1}), $w<-1$ and $w\to -1$ when
$t\to +\infty$.
In this model, there does not occur the Little Rip.
The inertial force in (\ref{i1}) generated by the expansion of the universe
is finite since the magnitudes of $H$ and $\dot H$ are bounded in the model 
(\ref{i6}).
Therefore the Little Rip does not occur although the magnitudes of $H$ 
and $\dot H$ become larger and larger in the model (\ref{i6}).

For $t \rightarrow \infty$, Eq.~(\ref{i7}) gives the asymptotic behavior of $w$
to be
\be
\label{wn1}
w \sim -1 - \frac{2\lambda H_1^{(\mathrm{II})}
\e^{- \lambda t}}{3\left(H_0^{(\mathrm{II})}\right)^2} \, ,
\ee
which is identical with (\ref{i4}) if we replace
${\lambda H_1^{(\mathrm{II})}}/{H_0^{(\mathrm{II})}}$
with $\lambda$.

By choosing $\alpha$ in (\ref{A5}) as
\be
\label{lr6}
\alpha (t) = \alpha_0^{(\mathrm{II})} \e^{-\lambda t}\, ,
\ee
with a constant $\alpha_0^{(\mathrm{II})}$,
we find $\omega(\phi)$ and $\eta(\chi)$ in (\ref{A5}) as follows
\bea
\label{lr7}
\omega (\phi) &=& \frac{2}{\kappa^2} \left( 
\sqrt{\left(\alpha_0^{(\mathrm{II})}\right)^2
+ \lambda^2 \left(H_1^{(\mathrm{II})}\right)^2}
 - \lambda H_1^{(\mathrm{II})} \right) \e^{-\lambda \phi}\, ,
\nn 
\eta (\chi) &=& - \frac{2}{\kappa^2} 
\sqrt{\left(\alpha_0^{(\mathrm{II})}\right)^2
+ \lambda^2 \left(H_1^{(\mathrm{II})}\right)^2}\,
\e^{-\lambda \chi} \, .
\eea
Using (\ref{A6}), one gets
\begin{align}
\label{lr8}
\tilde f(\phi,\chi) =& H_0^{(\mathrm{II})} - \frac{1}{\lambda}
\left\{ - \left( \sqrt{\left(\alpha_0^{(\mathrm{II})}\right)^2
+ \lambda^2 \left(H_1^{(\mathrm{II})}\right)^2} - \lambda H_1^{(\mathrm{II})} 
\right) \e^{- \lambda \phi} \right. \nn 
& \left. + \sqrt{\left(\alpha_0^{(\mathrm{II})}\right)^2
+ \lambda^2 \left(H_1^{(\mathrm{II})}\right)^2}\,\e^{-\lambda
\chi}\right\}\, ,
\end{align}
and the potential (\ref{A8}) is given by
\begin{align}
\label{lr9}
V(\phi,\chi) =& \frac{1}{\kappa^2} \left\{ \frac{3}{\lambda^2} \left\{ 
H_0^{(\mathrm{II})}
\lambda
+ \left( \sqrt{\left(\alpha_0^{(\mathrm{II})}\right)^2 + \lambda^2 
\left(H_1^{(\mathrm{II})}\right)^2}
 - \lambda H_1^{(\mathrm{II})} \right) \e^{- \lambda \phi} \right.\right. \nn
& \left. - \sqrt{\left(\alpha_0^{(\mathrm{II})}\right)^2
+ \lambda^2 \left(H_1^{(\mathrm{II})}\right)^2}\,\e^{-\lambda
\chi}\right\}^2 \nn
& \left. - \left( \sqrt{\left(\alpha_0^{(\mathrm{II})}\right)^2
+ \lambda^2 \left(H_1^{(\mathrm{II})}\right)^2} - \lambda H_1^{(\mathrm{II})} 
\right) \e^{-\lambda \phi}
+ \sqrt{\left(\alpha_0^{(\mathrm{II})}\right)^2
+ \lambda^2 \left(H_1^{(\mathrm{II})}\right)^2}\,\e^{-\lambda \chi}\right\}\, ,
\end{align}

For the model (\ref{i6}) with (\ref{lr6}), the eigenvalues (\ref{AAA1}) are
given by
\be
\label{lr9b}
M_\phi = M_\chi = - 3 + \frac{\lambda}{H}\, ,\quad
M_{\tilde Y} = - 3 - \frac{H_1^{(\mathrm{II})} \lambda 
\e^{-\lambda t}}{H^2} \, .
\ee
Therefore as long as $3 > \frac{\lambda}{H_0^{(\mathrm{II})}}$, the solution 
(\ref{i6}) is stable.

Since the solution can be unstable if $3 < \frac{\lambda}{H_0^{(\mathrm{II})}}$, 
we again consider the possibility that the universe could evolve to 
the de Sitter space-time.
For the potential (\ref{lr9}), there is an extremum when
\be
\label{stbl3}
 - \left( \sqrt{\left(\alpha_0^{(\mathrm{II})}\right)^2
+ \lambda^2 \left(H_1^{(\mathrm{II})}\right)^2}
 - \lambda H_1^{(\mathrm{II})} \right) \e^{\lambda \phi}
+ \sqrt{\left(\alpha_0^{(\mathrm{II})}\right)^2
+ \lambda^2 \left(H_1^{(\mathrm{II})}\right)^2}\,\e^{\lambda \chi} = -
\frac{\lambda^2}{6} + \lambda H_0^{(\mathrm{II})}\, ,
\ee
where the value of the potential $V(\phi,\chi)$ is given by
\be
\label{stbl4}
V(\phi,\chi) = \left( - \frac{\lambda^2}{12} + \lambda H_0^{(\mathrm{II})} 
\right)\, .
\ee
If $12 > \frac{\lambda}{H_0^{(\mathrm{II})}}$, which is consistent with
the condition $3 > \frac{\lambda}{H_0^{(\mathrm{II})}}$ 
that the solution (\ref{i6}) is stable,
$V(\phi,\chi)$ is positive and there is a solution corresponding to de Sitter 
space-time. Therefore there is a possibility that the universe could
evolve into  de Sitter space-time.
Note, however, the solution (\ref{stbl3}) corresponds to the minimum with
respect to both of $\phi$ and $\chi$ and therefore the de Sitter solution 
is not stable.

\subsection{Asymptotically de Sitter quintessence dark energy}

Let us consider the quintessence model where $w>-1$ but $w\to -1$ when 
$t\to +\infty$ as follows,
\be
\label{Ai6}
H = H_0^{(\mathrm{III})} + H_1^{(\mathrm{III})} \e^{-\lambda t}\, .
\ee
Here $H_0^{(\mathrm{III})}$, $H_1^{(\mathrm{III})}$, and $\lambda$ are
positive constants and we assume $H_0^{(\mathrm{III})}>H_1^{(\mathrm{III})}$
and $t>0$. The parameter $\lambda$ in (\ref{Ai6}) gives the same asymptotic
behavior of the EoS parameter $w$ ($w+1 \sim \e^{-\lambda t}$) in (\ref{lr1}).
Since the second term decreases when $t$ increases, the universe goes to
asymptotically de Sitter space-time.
Then from Eq.~(\ref{i2}), it follows
\be
\label{Ai7}
w = -1 + \frac{2\lambda H_1^{(\mathrm{III})} \e^{- \lambda t}}{3\left( 
H_0^{(\mathrm{III})}
  - H_1^{(\mathrm{III})} \e^{-\lambda t}
\right)^2} \, .
\ee
Hence, the EoS parameter is always larger than $-1$and $w\to -1$ when 
$t\to +\infty$. Therefore the universe is in non-phantom phase.
By choosing $\alpha$ in (\ref{A5}) as in (\ref{lr6}),
we find $\omega(\phi)$ and $\eta(\chi)$  (\ref{A5}) as follows
\begin{align}
\label{Alr7}
\omega (\phi) =& \frac{2}{\kappa^2} \left( 
\sqrt{\left(\alpha_0^{(\mathrm{II})}\right)^2
+ \lambda^2 \left(H_1^{(\mathrm{III})}\right)^2} 
+ \lambda H_1^{(\mathrm{III})} \right) \e^{-\lambda \phi}\, ,\nn
\eta (\chi) =& - \frac{2}{\kappa^2} 
\sqrt{\left(\alpha_0^{(\mathrm{II})}\right)^2
+ \lambda^2 \left(H_1^{(\mathrm{III})}\right)^2}\, \e^{-\lambda \chi} \, .
\end{align}
Using (\ref{A6}), one gets
\begin{align}
\label{Alr8}
\tilde f(\phi,\chi) =& H_0^{(\mathrm{III})} - \frac{1}{\lambda} \left\{
 - \left( \sqrt{\left(\alpha_0^{(\mathrm{II})}\right)^2
+ \lambda^2 \left(H_1^{(\mathrm{III})}\right)^2} + \lambda H_1^{(\mathrm{III})} 
\right) \e^{- \lambda \phi} \right. \nn 
& \left. + \sqrt{\left(\alpha_0^{(\mathrm{II})}\right)^2
+ \lambda^2 \left(H_1^{(\mathrm{III})}\right)^2}\,\e^{-\lambda
\chi}\right\}\, ,
\end{align}
and the potential (\ref{A8}) is given by
\begin{align}
\label{Alr9}
V(\phi,\chi) =& \frac{1}{\kappa^2} \left\{ \frac{3}{\lambda^2}
\left\{ H_0^{(\mathrm{III})} \lambda
+ \left( \sqrt{\left(\alpha_0^{(\mathrm{II})}\right)^2
+ \lambda^2 \left(H_1^{(\mathrm{III})}\right)^2} + \lambda H_1^{(\mathrm{III})} 
\right) \e^{- \lambda \phi} \right. \right. \nn
& \left. - \sqrt{\left(\alpha_0^{(\mathrm{II})}\right)^2
+ \lambda^2 \left(H_1^{(\mathrm{III})}\right)^2}\,\e^{-\lambda \chi}\right\}^2 \nn
& \left. - \left( \sqrt{\left(\alpha_0^{(\mathrm{II})}\right)^2
+ \lambda^2 \left(H_1^{(\mathrm{III})}\right)^2} + \lambda H_1^{(\mathrm{III})} 
\right) \e^{-\lambda \phi} + \sqrt{\left(\alpha_0^{(\mathrm{II})}\right)^2
+ \lambda^2 \left(H_1^{(\mathrm{III})}\right)^2}\,\e^{-\lambda \chi}\right\}
\, ,
\end{align}

For the model (\ref{Ai6}) with (\ref{lr6}), the eigenvalues (\ref{AAA1}) are
given by
\be
\label{AAAA2}
M_\phi = M_\chi = - 3 + \frac{\lambda}{H}\, ,\quad
M_{\tilde Y} = - 3 + \frac{H_1^{(\mathrm{III})} \lambda \e^{-\lambda t}}{H^2} 
\, .
\ee
If $3 > \frac{\lambda}{H}$ and $3 > \frac{H_1^{(\mathrm{III})} \lambda 
\e^{-\lambda t}}{H^2}$, the solution is stable.

Since the solution is unstable if $3 < \frac{\lambda}{H}$ or 
$3 < \frac{H_1^{(\mathrm{III})} \lambda \e^{-\lambda t}}{H^2}$, there is 
a possibility that the universe could evolve to the de Sitter space-time 
as in the model (\ref{lr9}).
For the potential (\ref{Alr9}), there is an extremum when
\be
\label{stbl5}
 - \left( \sqrt{\left(\alpha_0^{(\mathrm{II})}\right)^2
+ \lambda^2 \left(H_1^{(\mathrm{III})}\right)^2}
+ \lambda H_1^{(\mathrm{III})} \right) \e^{\lambda \phi}
+ \sqrt{\left(\alpha_0^{(\mathrm{II})}\right)^2
+ \lambda^2 \left(H_1^{(\mathrm{III})}\right)^2}\,\e^{\lambda \chi}
= - \frac{\lambda^2}{6} + \lambda H_0^{(\mathrm{III})}\, ,
\ee
where the value of the potential $V(\phi,\chi)$ is identical with that in
(\ref{stbl4}) and positive if $12 H_0^{(\mathrm{III})} > \lambda$.
The solution (\ref{stbl5}) is not stable, again.

\subsection{A realistic model unifying inflation with Little Rip dark energy 
era}

As one more  example, we consider the realistic model which contains the
inflation at $t\to - \infty$, phantom crossing at
$t=0$, and the Little Rip when $t\to \infty$:
\be
\label{ts1}
H = H_0^{(\mathrm{IV})} \cosh \lambda t\, .
\ee
Here $H_0^{(\mathrm{IV})}$ and $\lambda$ are positive constants.
The parameter $\lambda$ in (\ref{ts1}) gives the same asymptotic
behavior of the EoS parameter $w$ ($w+1 \sim \e^{-\lambda \left|t\right|}$) 
in (\ref{lr1}) for large $\left|t\right|$, that is, $w\to -1$ 
when $\left|t\right|\to \infty$. 
Since there are accelerating expansions when $t\to \pm\infty$, 
the one when $t\to -\infty$ can be identified with the inflation 
and the one when $t\to -\infty$ with the late time accelerating expansion. 
We should note that $H$ diverges in the limit of $t\to - \infty$, 
in addition to the limit of $t\to \infty$. This could tell that the classical 
description of the space-time by (\ref{ts1}) could become invalid in the limit 
of $t\to -\infty$ as expected usually and the quantum effect like the creation of the 
universe could be relevant. Therefore we should not consider 
seriously the limit of $t\to -\infty$ 
in the expression of (\ref{ts1}). 
Since $\dot H = H_0^{(\mathrm{IV})} \lambda \sinh \lambda t$,
we find $\dot H<0$ when $t<0$, that is, the universe is in non-phantom phase
and $\dot H>0$ when $t>0$, that is, the universe is in phantom phase.
There occurs the phantom crossing at $t=0$.
Therefore the present universe corresponds to $t\sim 0$.
When $\lambda t \gg 1$, we find that the Hubble rate $H$ behaves as
$H \sim \frac{H_0^{(\mathrm{IV})}}{2} \e^{\lambda t}$
and therefore there occurs the Little Rip.
The EoS parameter $w$ is now given by
\be
\label{ts3b}
w = - 1 - \frac{2\lambda \sinh \lambda t}{2 H_0^{(\mathrm{IV})} 
\cosh^2 \lambda t}\, .
\ee
Hence, $w<-1$ when $t>0$ and $w>-1$ when $t<0$. In the limit $t\to \pm \infty$,
$w\to -1$.
Thus when $t\to -\infty$, there occurs the accelerating expansion, which may
correspond to the inflation in the early universe.
When $w=-\frac{1}{3}$, that is,
\be
\label{ts3c}
\frac{\lambda \sinh \lambda t}{H_0^{(\mathrm{IV})} \cosh^2 \lambda t} = - 1 
\, ,
\ee
there occurs the transition between non-accelerating expansion and accelerating
expansion.
There are two negative solutions in (\ref{ts3c}) in general. Let us denote the
solution as $t_i$ and $t_l$ and assume $t_i<t_l<0$. Then $t=t_i$ corresponds 
to the end of inflation and $t=t_l$ to the transition from the non-accelerating 
expansion to the late accelerating expansion in the present universe.
More explicitly
\be
\label{ts3d}
\sinh \lambda t_i = - \frac{\lambda}{2H_0^{(\mathrm{IV})}}
 - \sqrt{\left( \frac{\lambda}{2H_0^{(\mathrm{IV})}} \right)^2 - 1}\, ,\quad
\sinh \lambda t_l = - \frac{\lambda}{2H_0^{(\mathrm{IV})}}
+ \sqrt{\left( \frac{\lambda}{2H_0^{(\mathrm{IV})}} \right)^2 - 1}\, .
\ee
Let the present universe corresponds to $t=t_\mathrm{present}$.
Since $t_i - t_\mathrm{present} = 137\times 10^8\sim
1/H_\mathrm{present} = 138\times 10^8$ years
($H_\mathrm{present} \sim 70 \mathrm{km}/\mathrm{s}\,\mathrm{Mpc}$), if we
assume $t_\mathrm{present}=0$,
however, Eq.~(\ref{ts3c}) does not have a solution for $t_i$. Then we may
assume $t_\mathrm{present} > 0$,
that is, the present universe is after the phantom crossing. In this case in
principle, one can solve (\ref{ts3c}) with respect to $\lambda$.
Then we can obtain the value of $t_l$ from the second equation in (\ref{ts3d}).
Roughly one can expect
the magnitude of the value could be $t_l \sim 50\times 10^8$ years and
therefore the realistic cosmology follows.

Eq.~(\ref{Acc}) shows that there occurs the rip between the earth and the sun 
when $a_\mathrm{e} \sim  \left(H_0^{(\mathrm{IV})}\right)^2
\e^{2 \lambda t}/4> a_\mathrm{g}$, which tells $\e^{2\lambda t} = 8.35\times 
10^{19}$ or $\lambda t = 22.9$.
If $\lambda = \mathcal{O}\left(10^{-10}\right)\, \mathrm{years}^{-1}$
as in (\ref{RRR1}), $t\sim 10^{11}\, \mathrm{years}$.

For the model (\ref{ts1}), by choosing $\alpha$ in (\ref{A5}) to be a constant
$\alpha = H_0^{(\mathrm{IV})} \lambda$, one finds
\be
\label{ts4}
\omega(\phi) = \frac{2H_0^{(\mathrm{IV})}\lambda}{\kappa^2}\e^{-\lambda \phi}
\, ,\quad
\eta(\chi) = - \frac{2H_0^{(\mathrm{IV})}\lambda}{\kappa^2} \cosh \lambda \chi 
\, .
\ee
Using (\ref{A6}), it follows
\be
\label{ts5}
\tilde f(\phi,\chi) = H_0^{(\mathrm{IV})} \left( \e^{-\lambda \phi}
+ \sinh \lambda \chi\right) \, ,
\ee
and the potential in (\ref{A8}) is given by
\be
\label{ts6}
V(\phi,\chi) = \frac{1}{\kappa^2} \left\{ \left(H_0^{(\mathrm{IV})}\right)^2 
\left( \e^{-\lambda \phi} + \sinh \lambda \chi
\right)^2 - H_0^{(\mathrm{IV})} \lambda \e^{-\lambda \phi} 
+ H_0^{(\mathrm{IV})} \lambda \cosh \lambda \chi \right\}\, .
\ee

For the model (\ref{ts4}) with $\alpha = H_0^{(\mathrm{IV})} \lambda$, the 
eigenvalues (\ref{AAA1}) are given by
\be
\label{ts6b}
M_\phi = - 3 + \frac{\lambda}{H_0^{(\mathrm{IV})} \cosh \lambda t}\, , \quad
M_\chi = M_{\tilde Y} = - 3 - \frac{\lambda \sinh \lambda t}
{H_0^{(\mathrm{IV})} \cosh^2 \lambda t}\, .
\ee
Therefore if $3>\frac{\lambda}{H_0^{(\mathrm{IV})}}$, all the eigenvalues are 
negative and therefore the solution is stable.

For the potential (\ref{ts6}), there is no extremum and therefore there does
not exist the solution corresponding to the de Sitter space-time.

Thus, we  constructed scalar models which describe the cosmological solutions
with and without Little Rip and investigated the
(in)stability of the solutions. We  also investigated the existence of the
solution describing de Sitter space-time and the stability of the de Sitter
solution when it exists as well as possible transition of Little Rip cosmology 
to de Sitter one. The results are summarized in Table \ref{table1}.

\begin{table}
\caption{The (in)stability of the solutions in the models}
\label{table1}
\begin{center}
\begin{tabular}
{|c|c|c|c|}
\hline
{\footnotesize Models} & {\footnotesize Stability of the reconstructed 
solution}
& {\footnotesize Existence of de Sitter solution} & {\footnotesize Stability of 
de Sitter solution} \\
\hline
(\ref{lr3}) & stable & no & $-$ \\
\hline
(\ref{lr7}) & stable if $3 > \frac{\lambda}{H_0^{(\mathrm{II})}}$ &
yes if $12 > \frac{\lambda}{H_0^{(\mathrm{II})}}$ & unstable \\
\hline
(\ref{Alr7}) & stable if $3 > \frac{\lambda}{H}$ and $3 > 
\frac{H_1^{(\mathrm{III})} \lambda \e^{-\lambda t}}{H^2}$
& yes if $12 > \frac{\lambda}{H_0^{(\mathrm{III})}}$ & unstable \\
\hline
(\ref{ts4}) & stable if $3 > \frac{\lambda}{H_0^{(\mathrm{IV})}}$ & no & $-$ \\
\hline
\end{tabular}
\end{center}
\end{table}

\section{Reconstruction in terms of e-foldings and  solution flow \label{IV}}

It is often used the e-foldings $N$ variable instead of the cosmological time 
$t$ since the e-foldings is  related with the redshift $z$ by
$\e^{-N} = 1 + z$, which is directly observed.
Here we consider the reconstruction of the two scalar model
in terms of the e-foldings $N$.
We also investigate the flow of the solution by defining
dimensionless variables, which give the fixed points for some solutions.

\subsection{Reconstruction of two scalar model and (in)stability}

Let us consider  two scalar model again.
By using the e-foldings $N$ the FLRW equations (\ref{A2}) and the scalar
field equations (\ref{A9}) are rewritten as
\begin{gather}
\label{2s2a}
1=\frac{\kappa^2}{6}\omega(\phi)\,\phi'{^2}
+\frac{\kappa^2}{6}\eta(\chi)\,\chi'{^2}
+\frac{\kappa^2V(\phi,\chi)}{3H^2}\,, \\
\label{2s2b}
1+\frac{2H'}{3H}=-\frac{\kappa^2}{6}\omega(\phi)\,\phi'{^2}
 -\frac{\kappa^2}{6}\eta(\chi)\,\chi'{^2}
+\frac{\kappa^2V(\phi,\chi)}{3H^2}\,, \\
\label{2s2c}
0=\omega(\phi)\left[\phi''+\left(3+\frac{H'}{H}\right)\phi'\right]
+\frac{1}{2}\omega_{,\phi}(\phi)\,\phi'{^2}
+\frac{V_{,\phi}(\phi,\chi)}{H^2}\,, \\
\label{2s2d}
0=\eta(\chi)\left[\chi''+\left(3+\frac{H'}{H}\right)\chi'\right]
+\frac{1}{2}\eta_{,\chi}(\chi)\,\chi'{^2}
+\frac{V_{,\chi}(\phi,\chi)}{H^2}\,,
\end{gather}
where $'$ denotes the derivative with respect to the e-foldings $N\equiv\ln a$.
A new function $\tilde{\mathcal{F}}(\phi,\chi)$ is defined by
\begin{equation}
\label{2s3}
\tilde{\mathcal{F}}(\phi,\chi)=f_0\exp\left\lbrace-\frac{\kappa}{2}
\left[\int\mathrm{d}\phi\,\omega(\phi)
+\int\mathrm{d}\chi\,\eta(\chi)\right]\right\rbrace\,.
\end{equation}
Here $f_0$ is a dimensionless constant.
The constants of the integration in (\ref{2s3}) 
do not affect the solution. In this sense, we have a class of model with 
two parameters, which generates the same solution. 
Let
\begin{equation}
\label{2s5}
\mathcal{F}(N)\equiv\tilde{\mathcal{F}}\left(\frac{N}{\kappa},
\frac{N}{\kappa}\right)\,.
\end{equation}
Then if the functions $\omega(\phi)$ and $\eta(\chi)$ satisfy
the following relations
\begin{equation}
\label{2s6}
\omega\left(\frac{N}{\kappa}\right)+\eta\left(\frac{N}{\kappa}\right)
=-2\frac{\mathcal{F}'(N)}{\mathcal{F}(N)}\,,
\end{equation}
and if the potential $V(\phi,\chi)$ is given by
\begin{equation}
\label{2s4}
V(\phi,\chi)=\frac{3\tilde{\mathcal{F}}(\phi,\chi)^2}{\kappa^4}
\left[1-\frac{\omega(\phi)+\eta(\chi)}{6}\right]\,.
\end{equation}
a solution of $\phi$, $\chi$ and $H$ is given by
\begin{equation}
\label{2s7}
\phi=\chi=\frac{N}{\kappa}\,, \quad
H=\frac{\mathcal{F}(N)}{\kappa}\,.
\end{equation}
Then one can obtain a model which reproduces arbitrary
expansion history of the universe given by $H=\mathcal{F}(N)/\kappa$,
by choosing $\omega(\phi)$, $\eta(\chi)$ and $V(\phi,\chi)$
by \eqref{2s4} and \eqref{2s6}.

\subsection{Fixed points and flow of general solutions}

Note that the solution \eqref{2s7} is one of the solutions
in the model \eqref{A1} with \eqref{2s4} and \eqref{2s6}.
In order to consider the structure of the space of the solutions,
by defining dimensionless variables,
we investigate the flow of the general solutions.
Besides the solution \eqref{2s7}, in general,
there are other solutions including the one describing the de Sitter space-time
corresponding to the extrema of the potential $V(\phi,\chi)$,
where $\phi$ and $\chi$ are constant.
We choose the variables so that both of the solution \eqref{2s7}
and the de Sitter solution correspond to fixed points.

Let us introduce the dimensionless variables as follows:
\begin{equation}
\label{2s8}
X=\kappa\phi'\,, \quad
Y=\kappa\chi'\,, \quad
Z=\frac{\phi}{\chi}\,, \quad
W=\kappa\left(\phi-\chi\right)\,.
\end{equation}
Eqs.~\eqref{2s2c} and \eqref{2s2d} are rewritten as
\begin{align}
\label{2s9a}
\begin{split}
X'&=3\left(1-X\right)+\frac{1}{2}\omega(\phi(Z,W))\left(X-X^2\right)
+\frac{1}{2}\eta(\chi(Z,W))\left(X-Y^2\right) \\
&\mathrel{\phantom{=}}
+\frac{\omega_{,\phi}(\phi(Z,W))}{2\kappa\,\omega(\phi(Z,W))}
\frac{6\left(1-X^2\right)-\eta(\chi(Z,W))\left(Y^2-X^2\right)}
{6-\omega(\phi(Z,W))-\eta(\chi(Z,W))}\,,
\end{split} \\
\label{2s9b}
\begin{split}
Y'&=3\left(1-Y\right)+\frac{1}{2}\omega(\phi(Z,W))\left(Y-X^2\right)
+\frac{1}{2}\eta(\chi(Z,W))\left(Y-Y^2\right) \\
&\mathrel{\phantom{=}}
+\frac{\eta_{,\chi}(\chi(Z,W))}{2\kappa\,\eta(\chi(Z,W))}
\frac{6\left(1-Y^2\right)-\omega(\phi(Z,W))\left(X^2-Y^2\right)}
{6-\omega(\phi(Z,W))-\eta(\chi(Z,W))}\,,
\end{split} \\
\label{2s9c}
Z'&=-\frac{\left(X-YZ\right)\left(1-Z\right)}{W}\,, \\
\label{2s9d}
W'&=X-Y\,.
\end{align}
Now the Hubble rate $H$ is given by
\begin{equation}
\label{2s10}
H=\frac{\tilde{\mathcal{F}}(\phi(Z,W),\chi(Z,W))}{\kappa}
\sqrt{\frac{6-\omega(\phi(Z,W))-\eta(\chi(Z,W))}
{6-\omega(\phi(Z,W))\,X^2-\eta(\chi(Z,W))\,Y^2}}\,.
\end{equation}
In order for  the Hubble rate to be real,
the values of $X$, $Y$, $Z$ and $W$ are restricted to a region
\begin{equation}
\label{2s11}
\frac{6-\omega(\phi(Z,W))\,X^2-\eta(\chi(Z,W))\,Y^2}
{6-\omega(\phi(Z,W))-\eta(\chi(Z,W))}>0\,.
\end{equation}
When $\omega(\phi)$ and $\eta(\chi)$ satisfy \eqref{2s6},
this system has two fixed points as follows:
\begin{description}
\item[Point A] $(X,Y,Z,W)=(1,1,1,0)$ \\
Here the solution is given by \eqref{2s7}.
\item[Point B] $(X,Y,Z,W)=(\beta_1,\beta_1,1,0)$ \\
Besides Point A $(X,Y,Z,W)=(1,1,1,0)$, there could be another solution  
for Eqs.~(\ref{2s9a}), (\ref{2s9b}), (\ref{2s9c}), 
and (\ref{2s9d}).  
In order to show the existence of another solution, we now define $\beta(N)$ by
\begin{equation}
\label{2s12}
\beta(N)\equiv\beta_0+\beta_1N\,.
\end{equation}
Here $\beta_0$ and $\beta_1$ are dimensionless constants. 
We now assume that the following equation could be satisfied, 
\begin{equation}
\label{2s13}
\frac{\omega_{,\phi}(\beta(N)/\kappa)}{\kappa\,\omega(\beta(N)/\kappa)}
=\frac{\eta_{,\chi}(\beta(N)/\kappa)}{\kappa\,\eta(\beta(N)/\kappa)}
=-\frac{6}{1+\beta_1}\left[1+\frac{f'(\beta(N))}{3f(\beta(N))}\right]
\left[1-\frac{f'(\beta(N))}{3f(\beta(N))}\beta_1\right]\,.
\end{equation}
Then if there exist $\beta_0$ and $\beta_1$ which satisfy \eqref{2s13}, 
we find Eqs.~(\ref{2s9a}), (\ref{2s9b}), (\ref{2s9c}), 
and (\ref{2s9d}) is satisfied by the following solution: 
\begin{equation}
\label{2s14}
\phi=\chi=\frac{\beta(N)}{\kappa}\,, \quad
H=\frac{f(\beta(N))}{\kappa}\sqrt{\frac{3f(\beta(N))+f'(\beta(N))}
{3f(\beta(N))+\beta_1{^2}f'(\beta(N))}}\,.
\end{equation}
Especially when $\beta_1=0$, this point describes de Sitter space-time.
\end{description}

We now choose $\omega(\phi)$ and $\eta(\chi)$ as
\be
\label{2s15a}
\omega(\phi) =-\frac
{f'(\kappa\phi)-\sqrt{f'(\kappa\phi)^2+\alpha(\kappa\phi)^2}}
{f(\kappa\phi)}>0\,, \quad
\eta(\chi) =-\frac
{f'(\kappa\chi)+\sqrt{f'(\kappa\chi)^2+\alpha(\kappa\chi)^2}}
{f(\kappa\chi)}<0\,,
\ee
where $\alpha(N)$ is an arbitrary function.
If we choose $\alpha(N)=\tilde\alpha_0\mathcal{F}'(N)$
and $\mathcal{F}(N)$ being a monotonically increasing or decreasing function,
$\omega(\phi)$ and $\eta(\chi)$ are respectively given by
\begin{equation}
\label{2s16}
\omega(\phi)=-\left(1-\epsilon\sqrt{1+\tilde\alpha_0{^2}}\right)
\frac{f'(\kappa\phi)}{f(\kappa\phi)}\,, \quad
\eta(\chi)=-\left(1+\epsilon\sqrt{1+\tilde\alpha_0{^2}}\right)
\frac{f'(\kappa\chi)}{f(\kappa\chi)}\,.
\end{equation}
Here $\tilde\alpha_0$ is a dimensionless constant
and $\epsilon\equiv \mathcal{F}'(N)/\lvert \mathcal{F}'(N)\rvert$.
Then $V(\phi,\chi)$ has the following form:
\begin{equation}
\label{2s17}
\begin{split}
V(\phi,\chi)&=\frac{3}{\kappa^4}
f(\kappa\phi)^{1-\epsilon\sqrt{1+\tilde\alpha_0{^2}}}
f(\kappa\chi)^{1+\epsilon\sqrt{1+\tilde\alpha_0{^2}}} \\
&\mathrel{\phantom{=}}\times\left[1
+\frac{1-\epsilon\sqrt{1+\tilde\alpha_0{^2}}}{6}
\frac{f'(\kappa\phi)}{f(\kappa\phi)}
+\frac{1+\epsilon\sqrt{1+\tilde\alpha_0{^2}}}{6}
\frac{f'(\kappa\chi)}{f(\kappa\chi)}\right]\,.
\end{split}
\end{equation}

\subsubsection{Model with exponential growth}

\begin{figure}
\centering
\begin{tabular}{cc}
\includegraphics{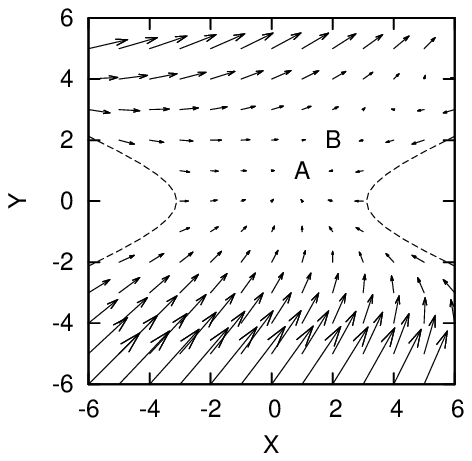} & \includegraphics{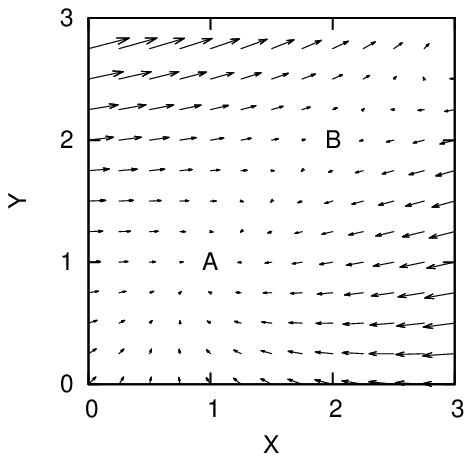}
\end{tabular}
\caption{%
Each vector denotes $(X'/50,Y'/50)$, which is independent of $Z$ and $W$.
The parameters are $\tilde\lambda=3/2$ and $\tilde\alpha_0=1$.
The point A is located in $(1,1)$, where the EoS parameter is $-2$.
The point B is located in $(2,2)$, where the EoS parameter is $-3$.}
\label{fig:2s1}
\end{figure}

\begin{figure}
\centering
\begin{tabular}{cc}
\includegraphics{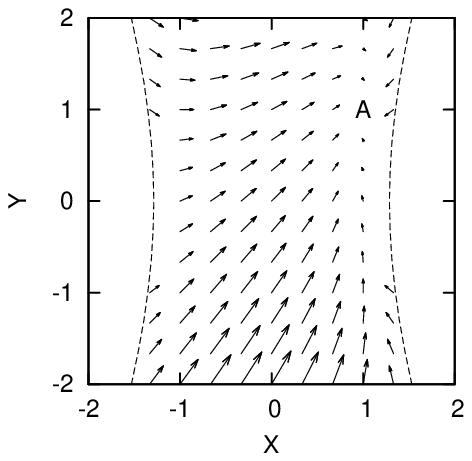} & \includegraphics{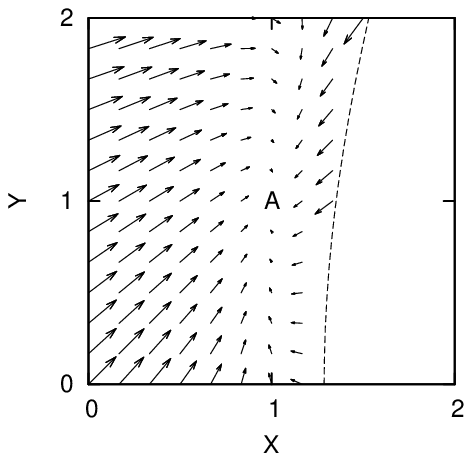}
\end{tabular}
\caption{%
Each vector denotes $(X'/20,Y'/20)$, which is independent of $Z$ and $W$.
The parameters are $\tilde\lambda=-3/2$ and $\tilde\alpha_0=1$.
The point A is located in $(1,1)$, where the EoS parameter is $0$.}
\label{fig:2s2}
\end{figure}

As an example, we consider
\begin{equation}
\label{2s18}
\mathcal{F}(N)=f_0\,\mathrm{e}^{\tilde\lambda N}\,,
\end{equation}
where $\tilde\lambda$ is a dimensionless constant.
Eq.~\eqref{2s17} has a solution $H(N)=f_0\,\mathrm{e}^{\tilde\lambda 
N}/\kappa$.
Then $\omega(\phi)$ and $\eta(\chi)$ are given by
\begin{equation}
\label{2s19}
\omega(\phi)=-\left(1-\epsilon\sqrt{1+\tilde\alpha_0{^2}}\right)\tilde\lambda\,, 
\quad
\eta(\chi)=-\left(1+\epsilon\sqrt{1+\tilde\alpha_0{^2}}\right)\tilde\lambda\,,
\end{equation}
and $V(\phi,\chi)$ is:
\begin{equation}
\label{2s20}
V(\phi,\chi)=\frac{3f_0{^2}}{\kappa^4}\left(1+\frac{\tilde\lambda}{3}\right)
\exp\left[\left(1-\epsilon\sqrt{1+\tilde\alpha_0{^2}}\right)\tilde\lambda\kappa\phi
+\left(1+\epsilon\sqrt{1+\tilde\alpha_0{^2}}\right)\tilde\lambda\kappa\chi\right]\,.
\end{equation}
The EoS parameter of the point A is
\begin{equation}
\label{2s21}
w=-1-\frac{2}{3}\tilde\lambda\,,
\end{equation}
which is independent of $N$. If $\tilde\lambda>0$,
the point B exists and is located in $(3/\tilde\lambda,3/\tilde\lambda,1,0)$.
Then the solution is given by
\begin{equation}
\label{2s22}
\phi=\chi=\frac{3\left(N-N_0\right)}{\tilde\lambda\kappa}\,, \quad
H=\frac{f_0}{\kappa}\sqrt{\frac{\tilde\lambda}{3}}
\,\mathrm{e}^{3\left(N-N_0\right)}\,.
\end{equation}
Here $N_0$ is an arbitrary constant.
The EoS parameter of this point is $-3$.
In this model, the dynamics of $X$ and $Y$ are independent of $Z$ and $W$.
Therefore, we consider a small fluctuation from each fixed point by
\begin{equation}
\label{2s23}
X(N)=X_0+\delta X(N)\,, \quad
Y(N)=Y_0+\delta Y(N)\,.
\end{equation}
Here $X_0$ and $Y_0$ are the values of $X$ and $Y$ in each fixed point.
Then \eqref{2s9a} and \eqref{2s9b} have the following form:
\begin{equation}
\label{2s24}
\frac{d}{dN}
\begin{pmatrix}
\delta X \\
\delta Y
\end{pmatrix}=
\begin{pmatrix}
 -3-\tilde\lambda+\left(1-\epsilon\sqrt{1
+\tilde\alpha_0{^2}}\right)\tilde\lambda X_0 &
\left(1+\epsilon\sqrt{1+\tilde\alpha_0{^2}}\right)\tilde\lambda Y_0 \\
\left(1-\epsilon\sqrt{1+\tilde\alpha_0{^2}}\right)\tilde\lambda X_0 &
 -3-\tilde\lambda+\left(1+\epsilon\sqrt{1+\tilde\alpha_0{^2}}\right)
\tilde\lambda Y_0
\end{pmatrix}
\begin{pmatrix}
\delta X \\
\delta Y
\end{pmatrix}\,.
\end{equation}
The eigenvalues of this matrix \eqref{2s24} are given by
\begin{equation}
\label{2s25}
\sigma_1=-3-\tilde\lambda\,, \quad
\sigma_2=-3-\tilde\lambda
+\left(1-\epsilon\sqrt{1+\tilde\alpha_0{^2}}\right)\tilde\lambda X_0
+\left(1+\epsilon\sqrt{1+\tilde\alpha_0{^2}}\right)\tilde\lambda Y_0\,.
\end{equation}
These indicate that the point A is stable if $-3<\tilde\lambda<3$
and unstable if $\tilde\lambda<-3$ or $\tilde\lambda>3$.
Similarly, the point B is stable if $\tilde\lambda>3$ and unstable if 
$0<\tilde\lambda<3$.
The dynamics of $X$ and $Y$ are shown
in \figurename~\ref{fig:2s1} and \figurename~\ref{fig:2s2}.

\subsubsection{Little Rip model}

\begin{figure}
\centering
\begin{tabular}{cc}
\includegraphics{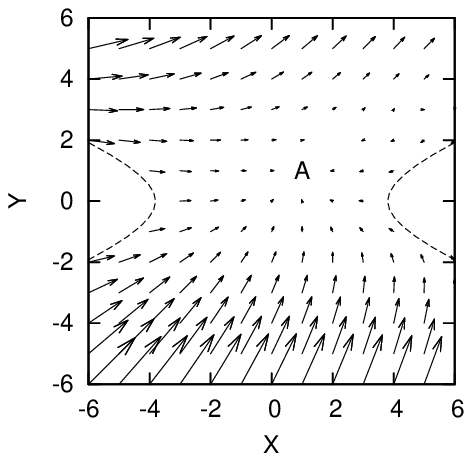} & \includegraphics{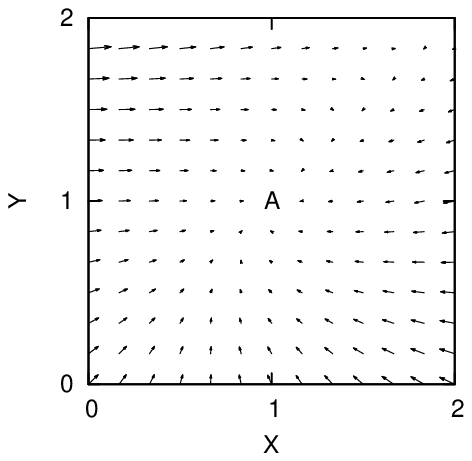}
\end{tabular}
\caption{%
Each vector denotes $(X'/50,Y'/50)$
with $Z=1$ and $W=0$ ($\phi=\chi=1/\kappa$).
The parameters are $\gamma=1$ and $\tilde\alpha_0=1$.
The point A is located in $(1,1)$,
which corresponds to the Little Rip universe.}
\label{fig:2s3}
\end{figure}

\begin{figure}
\centering
\begin{tabular}{cc}
\includegraphics{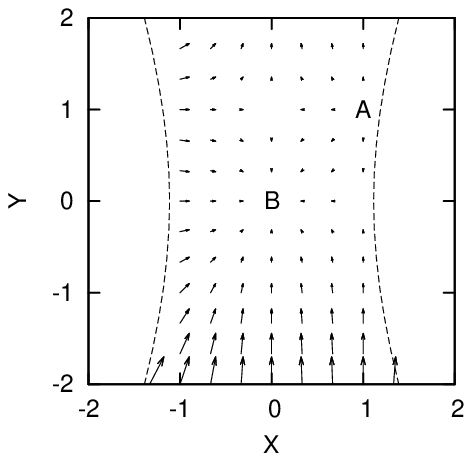} & \includegraphics{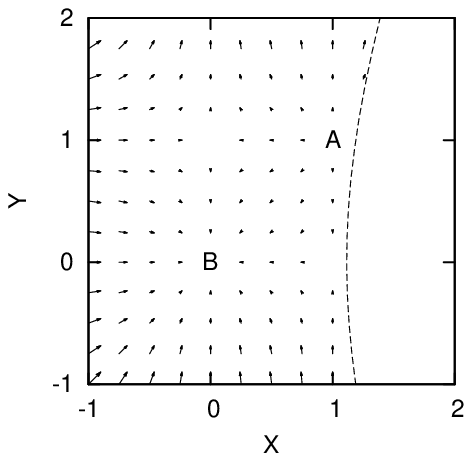}
\end{tabular}
\caption{%
Each vector denotes $(X'/20,Y'/20)$
with $Z=1$ and $W=0$ ($\phi=\chi=1/2\kappa$).
The parameters are $\gamma=-1$ and $\tilde\alpha_0=1$.
The point A is located in $(1,1)$.
The point B is located in $(0,0)$,
which corresponds to the de Sitter universe.}
\label{fig:2s4}
\end{figure}

We now consider another example
\begin{equation}
\label{2s26}
\mathcal{F}(N)=f_0N^\gamma\,,
\end{equation}
where $\gamma$ is a dimensionless constant.
We should note that the case $\gamma=1$ corresponds
to the Little Rip model (\ref{lr1}).
Then $\omega(\phi)$ and $\eta(\chi)$ are given by
\begin{equation}
\label{2s27}
\omega(\phi)=-\left(1-\epsilon\sqrt{1+\tilde\alpha_0{^2}}\right)
\frac{\gamma}{\kappa\phi}\,, \quad
\eta(\chi)=-\left(1+\epsilon\sqrt{1+\tilde\alpha_0{^2}}\right)
\frac{\gamma}{\kappa\chi}\,,
\end{equation}
and $V(\phi,\chi)$ has the following form:
\begin{equation}
\label{2s28}
\begin{split}
V(\phi,\chi)&=\frac{3f_0{^2}}{\kappa^4}
\left(\kappa\phi\right)^{1-\epsilon\sqrt{1+\tilde\alpha_0{^2}}}
\left(\kappa\chi\right)^{1+\epsilon\sqrt{1+\tilde\alpha_0{^2}}} \\
&\mathrel{\phantom{=}}\times\left[1
+\frac{1-\epsilon\sqrt{1+\tilde\alpha_0{^2}}}{6}
\frac{\gamma}{\kappa\phi}
+\frac{1+\epsilon\sqrt{1+\tilde\alpha_0{^2}}}{6}
\frac{\gamma}{\kappa\chi}\right]\,.
\end{split}
\end{equation}
The EoS parameter of the point A is
\begin{equation}
\label{2s29}
w=-1-\frac{2\gamma}{3N}\,,
\end{equation}
which becomes $-1$ when $N\to\infty$.
If $\gamma<1/2$, the point B exists and is located in $(0,0,1,0)$.
Then the solution is
\begin{equation}
\label{2s30}
\phi=\chi=\frac{1-2\gamma}{6\kappa}\,, \quad
H=\frac{f_0}{\sqrt{6}\kappa}\left(\frac{6}{1-2\gamma}\right)^{1/2-\gamma}\,.
\end{equation}
This point corresponds to the de Sitter space-time.
We consider a small fluctuation from the point B by
\begin{equation}
\label{2s31}
\phi(N)=\frac{1-2\gamma}{6\kappa}+\delta\phi(N)\,, \quad
\chi(N)=\frac{1-2\gamma}{6\kappa}+\delta\chi(N)\,.
\end{equation}
Then \eqref{2s9a} and \eqref{2s9b} have the following form:
\begin{align}
\label{2s32}
0= &\left(1-2\gamma\right)
\frac{d^2}{dN^2}
\begin{pmatrix}
\delta\phi \\
\delta\chi
\end{pmatrix} + 3 \frac{d}{dN}
\begin{pmatrix}
\delta\phi \\
\delta\chi
\end{pmatrix} \nn
& -18
\begin{pmatrix}
1-\left(1-\epsilon\sqrt{1+\tilde\alpha_0{^2}}\right)\gamma
& -\left(1+\epsilon\sqrt{1+\tilde\alpha_0{^2}}\right)\gamma \\
 -\left(1-\epsilon\sqrt{1+\tilde\alpha_0{^2}}\right)\gamma &
1-\left(1+\epsilon\sqrt{1+\tilde\alpha_0{^2}}\right)\gamma
\end{pmatrix}
\begin{pmatrix}
\delta\phi \\
\delta\chi
\end{pmatrix}\,.
\end{align}
The solution of this equation \eqref{2s32} is given by
\begin{equation}
\label{2s33}
\begin{pmatrix}
\delta\phi \\
\delta\chi
\end{pmatrix}=
\begin{pmatrix}
1+\epsilon\sqrt{1+\tilde\alpha_0{^2}} \\
 -1+\epsilon\sqrt{1+\tilde\alpha_0{^2}}
\end{pmatrix}
\frac{C_{1+}\,\mathrm{e}^{\sigma_{1+}N}
+C_{1-}\,\mathrm{e}^{\sigma_{1-}N}}{\kappa}
+ \begin{pmatrix}
1 \\
1
\end{pmatrix}
\frac{C_{2+}\,\mathrm{e}^{\sigma_{2+}N}
+C_{2-}\,\mathrm{e}^{\sigma_{2-}N}}{\kappa}\,.
\end{equation}
Here $C_{1\pm}$ and $C_{2\pm}$ are arbitrary constants
and $\sigma_{1\pm}$ and $\sigma_{2\pm}$ are given by
\begin{equation}
\label{2s34}
\sigma_{1\pm}=\frac{-3\pm3\sqrt{1+8\left(1-2\gamma\right)}}
{2\left(1-2\gamma\right)}\,, \quad
\sigma_{2\pm}=\frac{-3\pm3\sqrt{1+8\left(1-2\gamma\right){^2}}}
{2\left(1-2\gamma\right)}\,.
\end{equation}
Since $\sigma_{1+}$ and $\sigma_{1-}$ are positive, Eq.~(\ref{2s33}) 
tells that the fluctuation grows up when $N$ increases, which tells 
that the point B is unstable.
On the other hand, when $\phi$, $\chi\to\infty$,
$X$ and $Y$ in the solution behave as
\begin{equation}
X=1-X_1\,\mathrm{e}^{-3N}\,, \quad
Y=1-Y_1\,\mathrm{e}^{-3N}\,.
\end{equation}
Here $X_1$ and $Y_1$ are arbitrary constants.
This indicates that the values of $X$ and $Y$ approach to $(1,1)$
if $\phi$, $\chi\to\infty$ when $N\to\infty$.
Then the Hubble rate approaches to $f_0N^\gamma/\kappa$,
which corresponds to the Little Rip universe if $\gamma=1$.
The dynamics of $X$ and $Y$ are shown
in \figurename~\ref{fig:2s3} and \figurename~\ref{fig:2s4}.

\begin{figure}
\centering
\begin{tabular}{cc}
\includegraphics{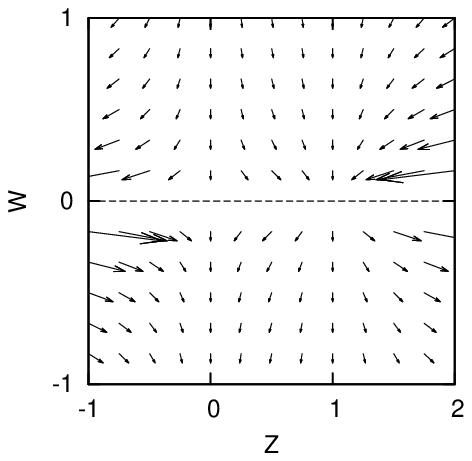} & \includegraphics{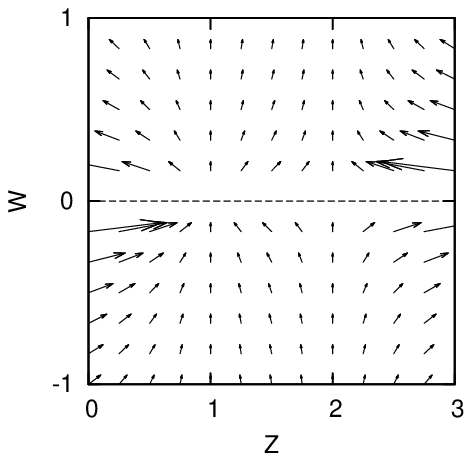} \\
(a) $X=0$, $Y=1$           & (b) $X=2$, $Y=1$           \\
\includegraphics{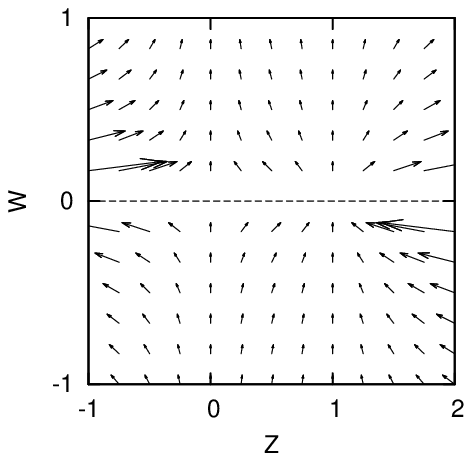} & \includegraphics{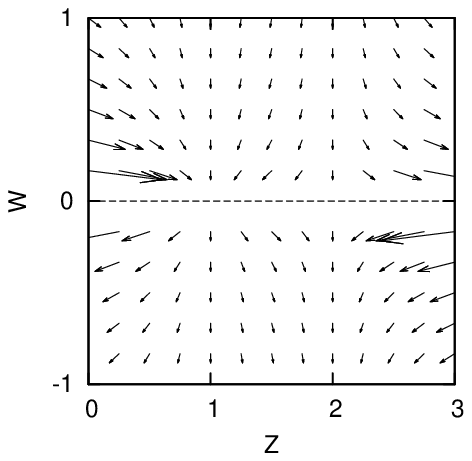} \\
(c) $X=0$, $Y=-1$          & (d) $X=-2$, $Y=-1$
\end{tabular}
\caption{%
Each vector denotes $(Z'/20,W'/20)$, which is independent of
the form of $\omega(\phi)$, $\eta(\chi)$ and $V(\phi,\chi)$.
The dynamics of $Z$ and $W$ are classified into four types
according to the values of $X$ and $Y$
as (a) $0<Y$, $X<Y$, (b) $0<Y<X$, (c) $0>Y$, $X>Y$ and (d) $0>Y>X$.
The fixed points are located in $(1,0)$.}
\label{fig:2s0}
\end{figure}

As mentioned above, the model (\ref{2s26}) with $\gamma=1$ corresponds
to the Little Rip model  (\ref{lr1}). In fact, Eq.~(\ref{lr1}) indicates that
\be
\label{AAAA1}
N = \frac{H_0^{(\mathrm{I})}}{\lambda} \e^{\lambda t}\, ,
\ee
and therefore
\be
\label{AAAA2B}
H = \lambda N\, .
\ee
By comparing (\ref{AAAA2B}) with (\ref{2s26}), we find $f_0 = \lambda$.
Then as written after (\ref{i4}), the parameter $f_0$ is bounded
as $2.37\times 10^{-3}\, \mathrm{Gyr}^{-1} \leq f_0 \leq 8.37\times
10^{-3}\, \mathrm{Gyr}^{-1}$ by the results of the Supernova Cosmology Project
\cite{Amanullah:2010vv}.
If we choose so that the present universe corresponds to $N=N_0$, we have
$f_0 N_0 \sim 70 \mathrm{km}/\mathrm{s}\,\mathrm{Mpc}$.

\subsection{The potential and the (in)stability}

Hence, we have completed the formulation of the reconstruction in terms
of the e-foldings. The e-foldings description is directly related with redshift
and therefore, with the cosmological observations.
We also investigated the flow for the solution, which shows the (in)stability
of the reconstructed solution obtained  in a large range.
Even if the solution is stable, when the stability region is small,
the evolution of the universe depends strongly  on the initial conditions.
If the initial condition is out of the range, the universe does not always
evolve to the solution obtained by the reconstruction.
On the other hand, if the stable region is large enough, even if the universe
started from an initial condition in a rather large region, the universe 
evolves to the solution obtained by the reconstruction.
The size of the stable region depend on the choice of $\alpha$ in (\ref{A5}). 
In other words, we may control the size of the stable region. 
Anyway for more quantitative analysis, we may need numerical calculation, which 
could be one of the future works. 

In case of the canonical scalar field as $\chi$, usually the scalar field rolls
down the potential. In case of the phantom scalar field as $\phi$, the scalar
field climbs up the potential. This phenomenon occurs due to the non-canonical
kinetic term. For the canonical scalar field $\varphi_c$, the field equation
has the form of $\nabla_t^2 \varphi_c = - V'(\phi)$ but for the phantom scalar
field $\varphi_p$, the sign of the kinetic term is changed, and we obtain
$\nabla_t^2 \varphi_p = V'(\phi)$ for a phantom scalar field. That is,
the sign of the ``force'' is effectively changed.
Then if there is an extremum in the potential, which corresponds to the
de Sitter space-time, the de Sitter solution can be stable if the extremum
corresponds to the minimum in $\chi$ and the maximum of $\phi$, otherwise
the de Sitter solution is unstable and the space-time could evolve to other
solution by a small perturbation.
The Rip generally occurs when the scalar fields climb up the potential and the
energy density of the scalar fields becomes larger and larger. If the scalar
fields arrive at the infinity of the potential in a finite time, there occurs
the Big Rip but if they arrive at the infinity when the cosmological time $t$
goes to infinity, there occurs the Little Rip. Then the solution corresponding
to the Big/Little Rip is stable, there is a path going to the infinity
of the potential and the path is minimum for the scalar field $\chi$. Otherwise
the universe may evolve to other solution.

\section{Discussion \label{VI}}

In summary, we gave a general formulation of reconstruction in two scalar model 
and investigated the stability of the solution. This formulation helped us to 
construct a model which has a stable cosmological solution describing
the phantom-divide crossing.
By using the formulation, we constructed non-singular phantom model which 
describes the cosmological solutions with and without Little Rip and 
investigated the (in)stability of the solutions. 
The existence of the solution describing (asymptotically) de Sitter space-time 
was also investigated and furthermore the stability of
the de Sitter solution when it exists as well as possible transition of
Little Rip cosmology to de Sitter one was investigated. We also considered
the reconstruction of the two scalar model
in terms of the e-foldings $N$ and investigated the flow of the solution by 
defining the dimensionless variables, which give the fixed points for some 
solutions.

Finally, let us make several remarks about the relation of the qualitative
behavior of the  Universe evolution and the shape of the scalar potential.

In case of the usual canonical scalar field as $\phi$ in (\ref{A1}),
when the field climbs up the potential, the kinetic energy decreases until
the kinetic energy vanishes. Even in case of the phantom field as $\chi$
in (\ref{A1}) with non-canonical kinetic term, the kinetic energy decreases
when the field climbs up the potential. In case of the phantom field, the
kinetic energy is unbounded below and therefore the absolute value of the
kinetic energy increases when the field climbs up the potential.
The Big Rip or Little Rip occurs when the potential goes to infinity.
If the potential tends to infinity in the finite future, the evolution
corresponds to the Big Rip but if the potential goes to infinity 
in the infinite future, the evolution corresponds to the
Little Rip. Then the necessary condition that the Big or Little Rip could 
occur is
\begin{enumerate}
\item The potential does not have maximum and it goes to infinity.
\item There is a path in the potential that the potential becomes infinite but
the kinetic energy of the canonical scalar field is vanishing.
\end{enumerate}
Since we identify the scalar field $\phi$ and $\chi$ with the cosmological
time, the second condition means $\omega(\phi)$ in (\ref{A1}) goes to zero 
when $\phi$ goes to infinity and therefore the phantom field $\chi$ dominates.
Conversely, if there is a maximum in the potential or there is no path in
the potential that the potential becomes infinite but
the kinetic energy of the canonical scalar field goes to zero,
there does not occur big rip nor Little Rip.

Let us suppose the case that there is a maximum in the potential.
If the fields stay near the potential maximum, the universe becomes
asymptotically de Sitter space-time.
If the fields go through the maximum and the potential decreases, the
kinetic energy of the canonical scalar field increases but the
absolute value of the kinetic energy of the phantom field decreases.
If the kinetic energy of the phantom field goes to zero, the canonical
field becomes dominant and the Universe could enter the non-phantom
(quintessence) phase and there might occur the deceleration phase in future.

In order to explain the accelerating expansion of the present universe, 
many kinds of models have been proposed. 
As clear even in this paper, only from the history of the expansion, 
we cannot find which model could be a real one. 
In order to distinguish between different models, the cosmological perturbation 
could be an important tool, which could be also one of the future works. 
Some works to distinguish between different models have been 
already reported in \cite{Copeland:2006wr}, for example.

\section*{Acknowledgments}

We are grateful to M.~Chaichian, P.~Frampton, V.~Mukhanov, M.~Sasaki,
and R.~Scherrer for related discussions.
This research has been supported in part
by MEC (Spain) project FIS2006-02842 and AGAUR(Catalonia) 2009SGR-994 (SDO),
by Global COE Program of Nagoya University (G07)
provided by the Ministry of Education, Culture, Sports, Science \&
Technology and by the JSPS Grant-in-Aid for Scientific Research (S) \# 22224003
and (C) \# 23540296 (SN).

\bibliographystyle{mdpi}
\makeatletter
\renewcommand\@biblabel[1]{#1. }
\makeatother

\end{document}